\documentclass[preprint,prd,amsmath,amssymb]{revtex4}
\usepackage{graphicx}
\usepackage{subfigure}
\usepackage{dcolumn}

\newcommand{\be}{\begin{equation}}
\newcommand{\ee}{\end{equation}}

\newcommand{\bea}{\begin{eqnarray}}
\newcommand{\eea}{\end{eqnarray}}
\newcommand{\bml}{\begin{multline}}
\newcommand{\eml}{\end{multline}}

\newcommand{\LE}{{\cal E}}

\newcommand{\LH}{{\cal H}}

\newcommand{\nn}{\nonumber}
\newcommand{\bm}[1]{\mbox{\boldmath $#1$}}

\newcommand{\hbm}[1]{\hat{\mbox{\boldmath $#1$}}}

\newcommand{\ord}{{\cal O}}

\newcommand{\boldx}{\boldsymbol{x}}
\newcommand{\eqn}[1]{Eq.~(\ref{#1})}

\newcommand{\eqns}[2]{Eqs.~(\ref{#1}) and (\ref{#2})}

\newcommand{\fig}[1]{Fig.~\ref{#1}}
\newcommand{\figs}[2]{Figs.~\ref{#1} and~\ref{#2}}
\newcommand{\figss}[2]{Figs.~\ref{#1}---\ref{#2}}

\newcommand{\tab}[1]{Table~\ref{#1}}

\newcommand{\sect}[1]{Section~\ref{#1}}

\newcommand{\rcite}[1]{Ref.~\onlinecite{#1}}

\begin{document}

\title{The Large $N$ Glueball Mass Spectrum in 2+1 Dimensions}

\author{Jesse Carlsson}
\email{j.carlsson@physics.unimelb.edu.au}
\author{Bruce H.~J.~McKellar}
\email{b.mckellar@physics.unimelb.edu.au}
\affiliation{School of Physics, The University of Melbourne}
\date{\today}
\begin{abstract}
   In this paper we explore the large $N$ limit of the glueball mass
spectrum for 2+1 dimensional pure gauge theory. We employ Hamiltonian
lattice gauge theory (LGT) and analytic variational techniques 
to calculate glueball masses for finite values of $N\le 25$. The
results are then accurately extrapolated to infinite $N$. An empirical
observation is discussed which links the infinite $N$ glueball
spectrum to a simple oscillator model.  
\end{abstract}
   
\maketitle

\section{Introduction}

In this paper we study the glueball mass spectrum for large $N$ pure
gauge theory (no quarks) in 2+1 dimensions using Hamiltonian LGT. 
Our primary motivation lies in recent developments in string
theory where a direct connection between pure gauge theory in the large
$N$ limit and certain string theories has been conjectured. An independent
test of this conjecture is provided by large $N$ calculations of the
glueball mass spectrum in LGT. 
Impressive Monte Carlo calculations by
Teper~\cite{Teper:1998te} and Lucini and Teper~\cite{Lucini:2002wg,Lucini:2001ej} have progressed to the stage where
accurate values of glueball masses have been calculated up to $N=6$ in
2+1 dimensions. This
has allowed an extrapolation to the $N\rightarrow \infty$ limit. With the
analytic techniques of the Hamiltonian approach~\cite{thesis-paper1} it is possible to extend
these calculations to much larger $N$, allowing a more reliable
$N\rightarrow \infty$ extrapolation.

The outline of this paper is as follows. We start in \sect{largenbackground} with
a historical background, where the original motivations for large $N$
physics are discussed and the recent developments in string theory
introduced.
In \sect{extrapolation} we use the analytic variational approach
discussed in \rcite{thesis-paper1} to calculate glueball masses at
finite values of $N$ up to $N=25$ in some cases. These results are
then extrapolated to explore the large $N$ limit of the 
2+1 dimensional glueball mass spectrum. 
The $N\rightarrow \infty$ limits obtained here are 
compared to the results of the analogous Monte Carlo studies in the Lagrangian approach.

\section{Background}
\label{largenbackground}
The nonperturbative region of quantum chromodynamics (QCD) 
has been examined predominantly with
numerical techniques. While the numerical Monte Carlo simulations of 
Lagrangian LGT have made significant progress, there have been
few developments in analytic techniques. One analytic technique which has 
received considerable attention in recent years is the large $N$ limit. 

In 1974 't Hooft proposed a study of SU($N$) gauge theories in the
large $N$ limit rather than the physically interesting $N=3$ case~\cite{'tHooft:1974jz}. It
was hoped that the SU($N$) theory could be solved analytically and
would be, in some sense, close to SU(3). 
Indeed it was shown that the large $N$ theory simplifies
drastically~\cite{'tHooft:1974jz} but still captures at least some of the complexity of
the SU(3) theory~\cite{Witten:1979kh, Eguchi:1982nm}. 
Motivated by the fact that the QCD coupling constant $g^2$ is a
poor expansion parameter because its value depends on the energy scale of the process under consideration, t' Hooft considered an expansion in 
another dimensionless parameter of SU($N$) QCD, $1/N$. Based on an ingenious
organisation of Feynman diagrams, 't Hooft was able to show that in
the large $N$ limit, keeping the 't Hooft coupling $g^2 N$ fixed, only planar diagrams
remain. While the remaining planar theory can be solved exactly in
two dimensions, the three and four dimensional theories have not yet
been solved. 

Recent developments in string theories have attempted to address this
problem. In 1997 Maldacena conjectured a correspondence, in the
duality sense, between
superconformal field theories and string theory propagating in a
non-trivial geometry~\cite{Maldacena:1998re}.  The idea
behind duality is
that a single theory may have two (or more) descriptions with the
property that when one is strongly coupled the other is weakly
coupled. A technique for breaking the conformal
invariance and supersymmetry constraints of Maldacena's original
proposal was later developed by Witten~\cite{Witten:1998qj}. 
This led to the hope that
nonperturbative pure SU($N$) gauge theories could be described
analytically in 3+1 dimensions by their string theory dual. To develop
this hope into a concrete solution would require perturbative
expansions within the dual string theory. Such developments would seem to be some
way off.

In order to test the viability of using string theories 
to probe nonperturbative QCD independent
tests are needed. The glueball mass spectrum of QCD provides a perfect
laboratory. Recent Monte Carlo calculations have provided stable
estimates of glueball masses up to $N=6$~\cite{Teper:1998te,Lucini:2001ej} in 2+1 dimensions and up to $N=5$~\cite{Lucini:2001ej} in
3+1 dimensions. From these estimates an $N\rightarrow
\infty$ limit can be extrapolated in each case. Convincing evidence of 
$\ord(1/N^2)$ finite $N$ corrections have been demonstrated in 2+1 dimensions,
verifying a specific prediction of 't Hooft's $1/N$ expansion in the
quarkless case.  From the string theory side, present
calculations require a strong coupling limit ($N, g^2 N\rightarrow
\infty $) to be taken. In this limit the string theory reduces to classical supergravity and 
the calculation of glueball masses, $m(J^{PC})$, is straightforward. In this limit the 
glueball is represented by a dilaton
field whose mass can be extracted by solving the dilaton wave
equation~\cite{Csaki:1998qr,deMelloKoch:1998qs,Zyskin:1998tg}. 
Despite
the extreme approximations required surprising quantitative agreement
with the weak coupling results from LGT were obtained. Later it was
realised that, in 2+1 dimensions, lower mass states can be constructed from the graviton
field~\cite{Brower:1999nj}. 
With this approach the quantitative agreement with LGT is
lost but qualitative agreement in the form of the prediction 
\bea
m(0^{++}) < m(2^{++}) < m(1^{-+})
\eea   
still holds. The qualitative agreement between
string theories and LGT persists
in 3+1 dimensions~\cite{Brower:2000rp} with
 \bea
m(0^{++}) < m(2^{++}) < m(0^{-+}),
\eea  
for both theories. It appears that the identification of glueball states
remains a problem. A serious difficulty is the removal of unwanted states which have masses of the same magnitude as the glueballs. Recipes for removing these unwanted states have been
proposed~\cite{Russo:1998mm,Csaki:1999vb} but new parameters need to be
introduced.

The exploration of large $N$ gauge theories outside of string theory 
has not been restricted to
traditional Monte Carlo methods. Recent studies using the light-front 
Hamiltonian of transverse LGT have shown agreement with the Monte
Carlo calculations of Teper~\cite{Dalley:2000ye}. 
In this approach explicit calculations at
$N\rightarrow \infty$ are possible without the need for
extrapolation. Continuum calculations, without the use of string
theory, have also commenced. An impressive series of papers~\cite{Karabali:1996je,Karabali:1998wk,Karabali:1998yq}
has led to specific predictions for the string tension for all $N$ in
2+1 dimensions~\cite{Karabali:1998yq}. These predictions lie within $3\%$ of lattice
calculations up to $N=6$ beyond which LGT results are not available. A
recent review is available in \rcite{Nair:2002uj}.

\section{Extrapolation}
\label{extrapolation}

\subsection{Introduction}

It this section we calculate 2+1 dimensional SU($N$) glueball masses 
for various $N$ using the analytic variational technique of
\rcite{thesis-paper1}. Our intention is to explore the large $N$
limit of glueball mass spectrum by extrapolating the results obtained
to their $N\rightarrow \infty$ limit. Employing analytic methods we can
extend the calculation of the lowest energy massgaps to SU(25) on a
desktop computer. This should be compared to current results in the
Lagrangian approach which are limited, at present, to SU(6). With large values of $N$
available a more reliable extrapolation to the $N\rightarrow \infty$
limit is possible.

To simulate the ground, or (perturbed) vacuum, state with energy
$E_0$, we use the one plaquette trial state, 
\be
\begin{array}{c}\includegraphics{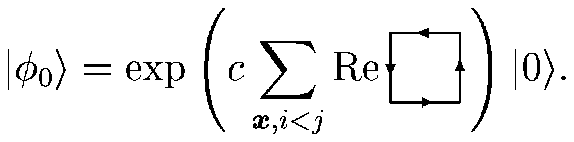}\end{array}
\label{oneplaquette}
\ee
Here, $|0\rangle $ is the strong coupling vacuum defined by
$\LE^\alpha_i(\bm{x})|0\rangle = 0$ for all $i$, $\bm{x}$ and $\alpha =
1,2,\ldots,N^2-1$. $\LE^\alpha_i(\bm{x})$ is the lattice
chromoelectric field on the directed link running 
from the lattice site labelled by $\bm{x}$ to that labelled by
$\bm{x}+a\hbm{i}$. Here $\hbm{i}$ is a unit vector in the $i$
direction. The directed square denotes the
traced 
ordered product of link 
operators, $U_i(\bm{x})$, around an elementary square, or plaquette, of the lattice,
\be
\begin{array}{c}\includegraphics{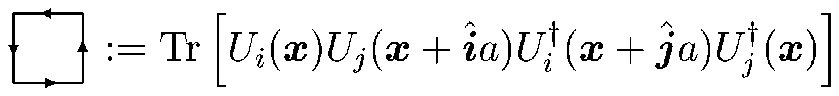}\end{array}
\ee
where $a$ is the lattice spacing.

As discussed in \rcite{thesis-paper1}, the precise form of $c(\beta)$,
where $\beta = N/g^2$, is unimportant in a scaling region. 
As it turns out, to calculate variational wave functions, by
minimising the vacuum energy density, 
for large dimension gauge groups is cumbersome. Numerical precision
becomes a factor in the minimisation of the energy density. This
problem is magnified in the calculation of tadpole improved results
using the iterative procedure employed in \rcite{thesis-paper1}. 
We thus abandon the use of variational wave functions here. Instead
we make use of the one plaquette wave function of \eqn{oneplaquette} and
define a simple dependence of $c(\beta)$. Here we choose $c(\beta)
=\beta$. 

Many Hamiltonians are available for LGT calculations~\cite{Carlsson:2001wp}. In
\rcite{thesis-paper1} we used Kogut-Susskind~\cite{Kogut:1975ag},
improved and tadpole improved Hamiltonians to calculate glueball
masses for $N\le 5$. Here we employ the simplest of these, the
Kogut-Susskind Hamiltonian, which is defined for pure SU($N$) gauge
theory with coupling $g^2$ on a lattice with
spacing $a$ by 
\bea
\LH &=& \frac{g^2}{2a}\sum_{\boldx,i}
\LE^\alpha_i(\boldx)^2 + \frac{2N}{a g^2} \sum_{\boldx, i<j} P_{ij}(\boldx),
\label{kogut-susskind}
\eea
where the plaquette operator is given by
\be
\begin{array}{c}\includegraphics{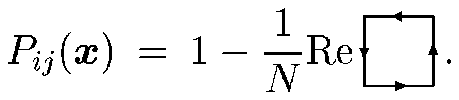}\end{array}
\label{plaquetteoperator}
\ee

To evaluate the matrix elements appearing in the glueball mass
calculation we use the SU($N$) generating functions derived in
\rcite{thesis-paper1},
\bea
G_{{\rm SU}(N)}(c,d) &=& \int_{{\rm SU}(N)} dU e^{c {\rm Tr} U + d {\rm
Tr} U^\dagger}\nn\\
&=&
\sum_{l=-\infty}^{\infty} \left(\frac{d}{c}\right)^{l N/2}
\det \left[ I_{l+j-i}\left(2\sqrt{cd}\right)\right]_{i\le i,j \le N}
\label{coolsum}
\eea
and
\bea
H_m(c,d) &=& \int_{{\rm SU}(N)} dU \exp\left[c ({\rm Tr} U + {\rm Tr}
U^\dagger)+ d{\rm Tr}(U^m)\right]\nn\\
&=& \sum_{l=-\infty}^{\infty}\det
\left[\sum_{k=0}^{\infty} \frac{d^k}{k!} I_{l+j-i+m k}(2 c)\right]_{1\le i,j\le N} .
\label{coolersum}
\eea
Here the quantities inside the determinant are to be interpreted as
the $(i,j)$-th entry of an $N\times N$ matrix. 

For the minimisation process, we use the same basis of rectangular states as in
\rcite{thesis-paper1}. With this basis the states $J^{++}$ and
$J^{--}$, with $J=0$ or 2, are accessible. The $J=2$ states require the use of a minimisation basis that contains no square states so that the excited states are invariant under rotations by $\pi$ but not $\pi/2$. The dependence of the
massgaps on the coupling does not change significantly as $N$ is
increased. We do however observe the appearance of what could possibly
be additional low $\beta$ scaling regions as $N$ is increased.


\subsection{Convergence with $\bm{l_{{\rm max}}}$}
\label{convwithl}

\begin{figure}
\centering
\subfigure[SU(4) $\Delta M^{++}$. ] 
                     {
                         \label{symconv}
                         \includegraphics[width=7cm]{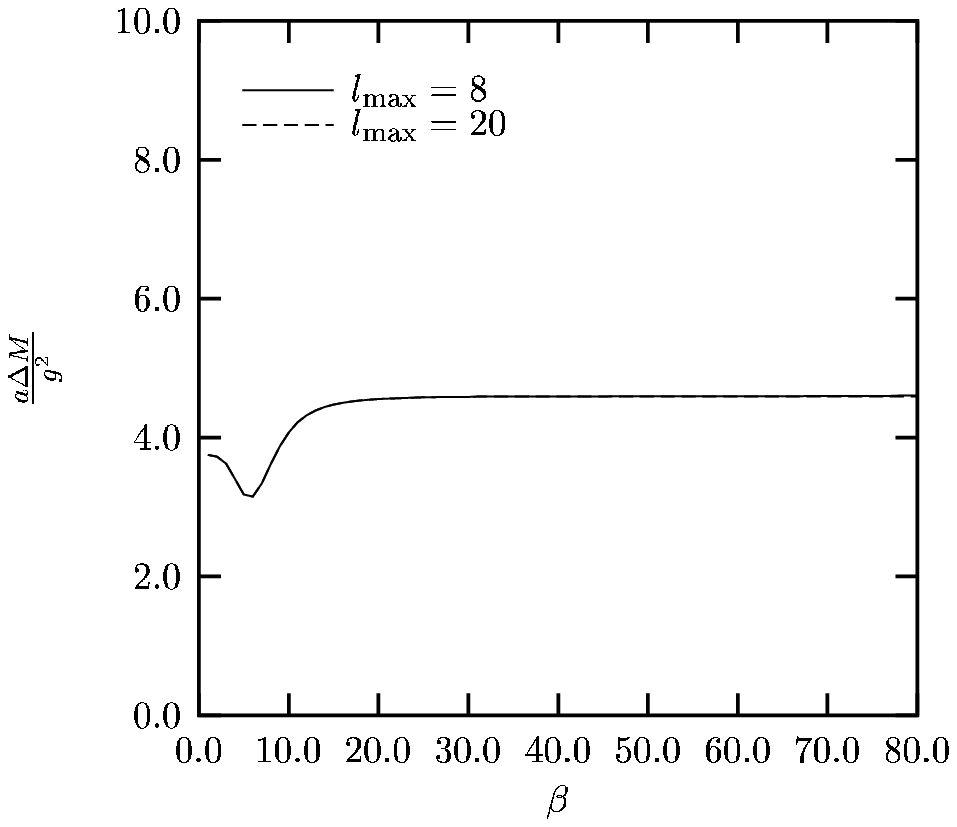}
                     }    \hspace{0.25cm}                     
\subfigure[SU(3) $\Delta M^{--}$.] 
                     {
                         \label{asymconv}
                       \includegraphics[width=7cm]{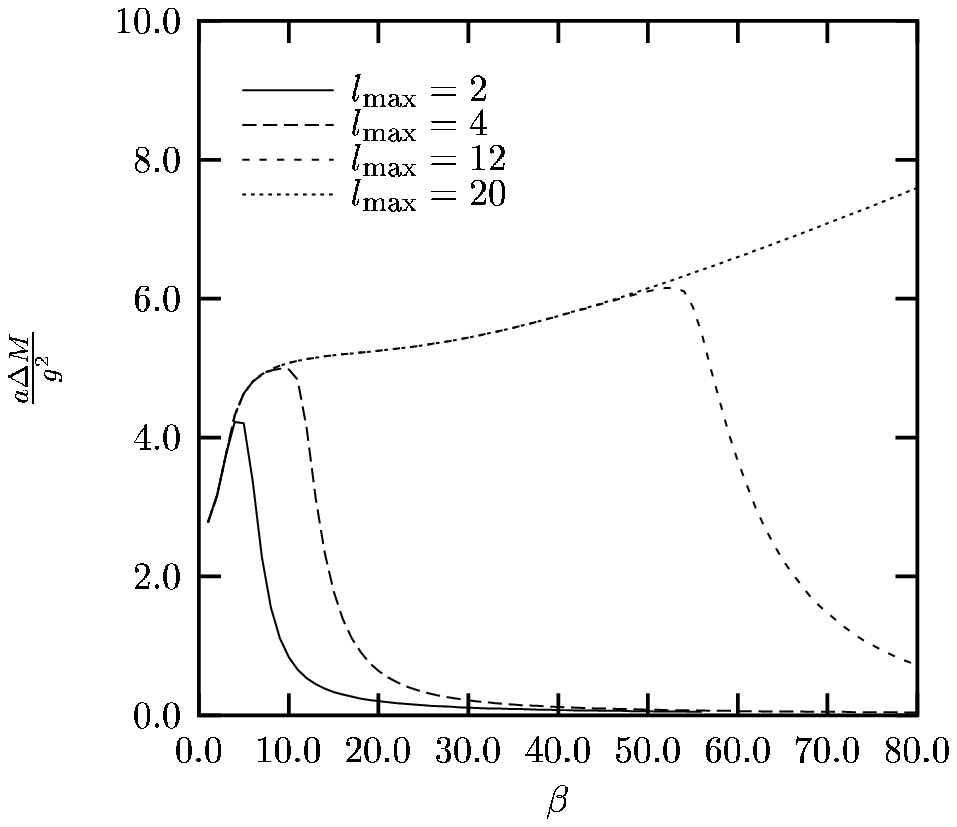}
                     }                  

\caption{Example spin 0 massgaps in units of $g^2/a$  demonstrating 
         different convergence properties when truncating the
         $l$-sum of \eqn{coolsum} at $l=\pm l_{\rm max}$.} 
\label{massgapconvergence}  
\end{figure}

In \rcite{thesis-paper1} the dependence of the
variational parameter on the truncation of the $l$-sum in \eqn{coolsum}
was considered. The calculation of the variational parameter depends
only on the plaquette expectation value. The dependence of the
massgaps on the truncation should also be explored as their
calculation incorporates additional matrix elements. We find that fast
convergence is achieved as $l_{\rm max}$ is increased for the $0^{++}$ massgaps. A typical example
is shown in \fig{symconv}, where the massgaps are almost
indistinguishable up to $\beta =80$ on the scale of the plot. For
$0^{--}$ the convergence is not as fast. A typical example is shown in
\fig{asymconv}. We find that in order to obtain convergence of
the antisymmetric massgaps up to $1/g^2 \approx 350$ for SU(25) we
need a truncation of $l_{\rm max}=30$.

\subsection{The small $\bm{\beta}$ minima}

The massgaps for $N\ge 2$, in units of $N g^2/a$ do not differ
significantly as functions of $1/g^2$. The lowest lying $0^{++}$
glueball masses have
the characteristic form shown in \fig{symconv} for each $N$. We observe a
minimum at small $\beta$ and a scaling plateau at
large $\beta$. 

In this section we examine the minima occurring in the lowest $0^{++}$
and $0^{--}$ glueball masses and consider the possibility that they
may correspond to a scaling region. It appears that for SU(3) this
region has been interpreted as a scaling region by some authors~\cite{Fang:2002ps,Chen:1995ca,ConradPhD}. From the scaling
arguments of \rcite{thesis-paper1}, by changing the functional
dependence of the parameter appearing in the vacuum state the scaling
region can be modified. In this way it is possible to choose the
variational parameter's dependence on $\beta$ such that the region in
which the mass gap takes its minimum is stretched over a large
$\beta $ interval. In the same way, the scaling plateaux occurring
for large $\beta $ may be compressed. With this in mind, it is
unclear which scaling region we should take to be the correct one.
For this reason we analyse both potential scaling regions. We start, 
in this section, with the small $\beta$ minima and consider the large
$\beta$ plateaux in the next section.

\begin{figure}
\centering

\includegraphics[width=10cm]{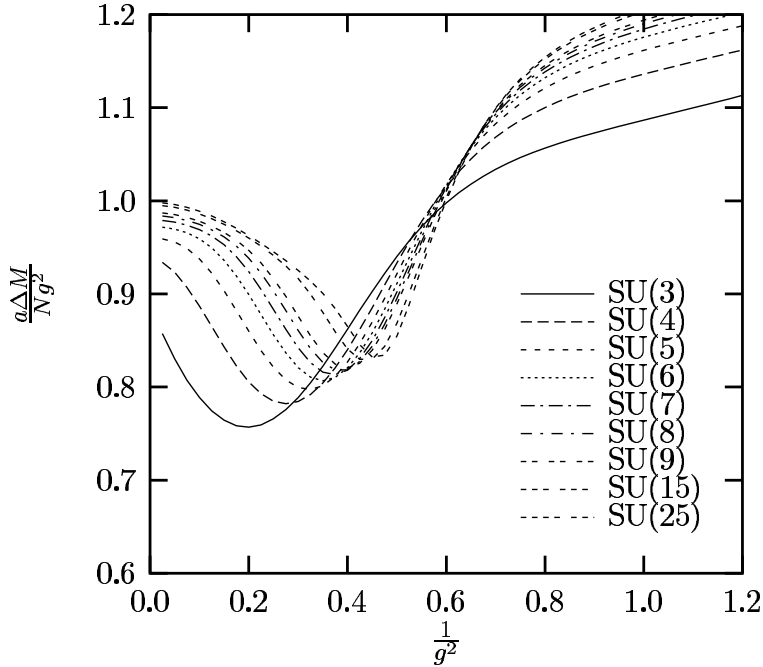}

\caption{$L_{max} = 4$  
SU($N$) lowest mass symmetric massgaps in 2+1 dimensions in units of
$g^2 N/a$ as functions of $1/g^2$. The $l$-sum of
\eqns{coolsum}{coolersum} truncated at $l_{{\rm max}}=5$. } 
\label{smallbetamin}  
\end{figure}

\fig{smallbetamin} shows the lowest lying SU($N$) $0^{++}$
glueball masses in 2+1 dimensions in units of $g^2 N/a$ as functions of
$1/g^2$ for various $N\in[3,25]$. It is apparent that the minima
depend only weakly on $N$. We see that the minima appear to approach a
finite limit from below as $N\rightarrow \infty$. On the scale of
\fig{smallbetamin} the minima corresponding to $N=15$ and
$N=25$ are barely distinguishable.

\begin{figure}
\centering
                       
\includegraphics[width=10cm]{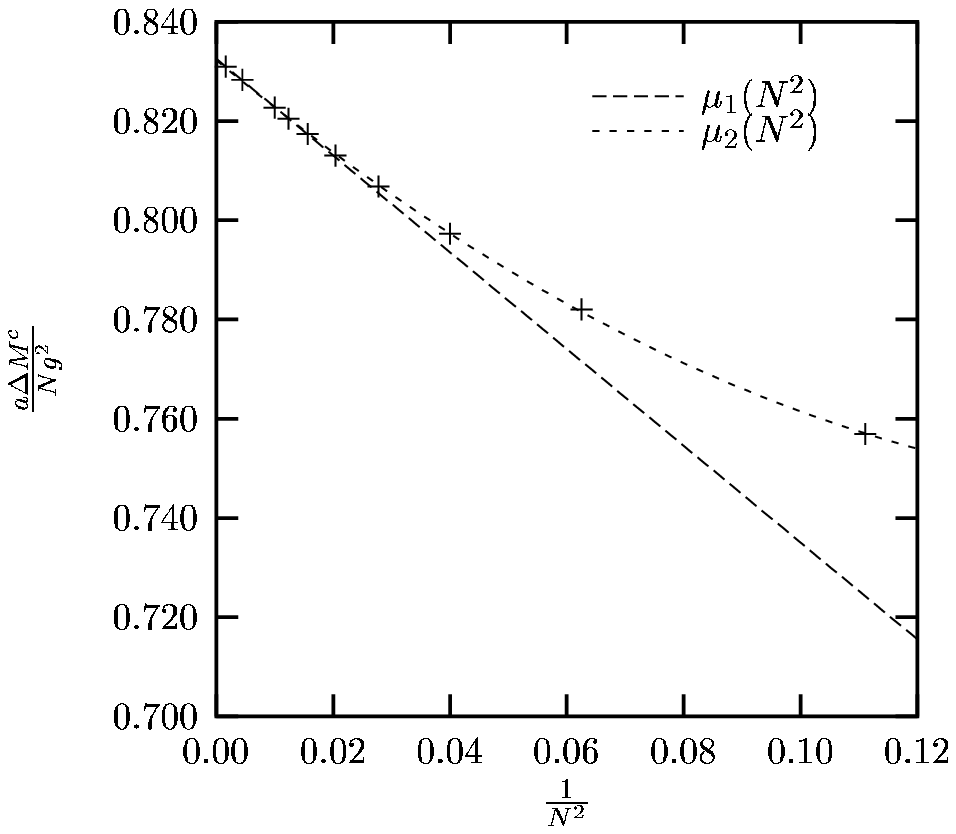}

\caption{ The continuum limit lowest mass symmetric 2+1 dimensional massgaps in units
of $N g^2/a$ as a
function of $1/N^2$ taken from the small $\beta$ minima. The dashed lines are fits to \eqns{smallbetafit1}{smallbetafit2}.}
\label{smallbetalargen}  
\end{figure}

Let us denote the minima in \fig{smallbetamin} by $a\Delta
M^c/(N g^2)$ and consider them as a functions of
$N$. Fitting $a\Delta
M^c/(N g^2)$ to the model 
\bea
\frac{a \Delta M^c}{N g^2} = \gamma_1 + \frac{\gamma_2}{N^{2 \gamma_3}} 
\eea
for $N \ge 8$ gives best fit parameters
\bea
\gamma_1  &=& 0.83262 \pm 0.00022  \nn\\
\gamma_2  &=& -0.97 \pm 0.13      \nn\\
\gamma_3  &=& 0.990 \pm 0.035 .
\label{fitone-1}
\eea
We fit on the data $N \ge 8$ to minimise pollution from next to
leading order corrections. 
If we assert that the power of the leading order correction for
finite $N$ must be an integer power of $1/N$, then from \eqn{fitone-1}
that order must be 2 in agreement with the predictions of large $N$ QCD. 
Fitting to a model with $1/N^2$ corrections for $N\ge 8$ gives
\bea
\mu_1(N^2) = 0.83256 \pm 0.00007 - \frac{0.9753 \pm 0.0072}{N^2}.
\label{smallbetafit1}
\eea 
We can go further and attempt to find the next to leading order
corrections by fitting to the model
\bea
\frac{a \Delta M^c}{N g^2} = \gamma_1 + \frac{\gamma_2}{N^{2}} +
\frac{\gamma_3}{N^{2 \gamma_4}}. 
\eea
The best fit parameters when fit on the whole data set are 
\bea
 \gamma_1 &=&      0.83287     \pm    0.00011   \\
\gamma_2 &=& -1.116      \pm    0.028     \\
\gamma_3 &=&  1.715     \pm     0.085     \\
\gamma_4 &=& 1.627      \pm    0.050   
\eea
which is in disagreement with integer power next to leading order
corrections. A good fit however is achieved to the model with $1/N^4$
next to leading order corrections:
\bea
\mu_2(N^2) &=& 0.83233 \pm 0.00018
 - \frac{0.986 \pm 0.010}{N^2} + \frac{2.772 \pm 0.090}{N^4}.
\label{smallbetafit2}
\eea
The precise locations of the minima, $a\Delta
M^c/(N g^2)$, are plotted as a funtion of $1/N^2$ in \fig{smallbetalargen}
along with the fitted models of \eqns{smallbetafit1}{smallbetafit2}.

From \eqn{smallbetafit1}, our best fit result for the
$N\rightarrow \infty$ limit of  $a\Delta M^c/(N g^2)$ is  $0.83256 \pm
0.00007$. The error here is purely statistical. We should, as
always, expect a significant systematic error attributable to our
choice of ground state and minimisation basis. This result may
be compared to the Monte Carlo calculation of Lucini and Teper who obtain the 
$N\rightarrow \infty$ limit of the lowest $0^{++}$ glueball mass as $0.8116\pm
0.0036$~\cite{Lucini:2002wg}. This is comparable to the result
presented here. The result of Lucini and Teper is based on a linear extrapolation to
the $N\rightarrow \infty$ limit of $2 \le N \le 6$ data. 
Their result differs significantly from the one presented here at the 
leading order finite $N$ corrections. The correction term  obtained by Lucini and Teper, 
$-(0.090\pm 0.028)/N^2$, has the same sign but is significantly smaller than the
one presented here. When our
data is fit on the range $3\le N \le 6$ (we don't obtain a small
$\beta$ minimum for SU(2)) our slope is halved but is still significantly
larger than that of Lucini and Teper. It should be pointed out that
Lucini and Teper's calculation was 
performed in the Lagrangian approach in which the coupling is the
so called Euclidean coupling $g_E$. The Euclidean coupling and the
Hamiltonian coupling, $g^2$, are equal up to order $g_E^2$
corrections. The precise relation between the couplings is 
only known for small $g_E^2$~\cite{Hamer:1996ub}, a case which does not apply for the
small $\beta $ minima. It appears that our simple calculation induces
a level crossing whereby the lowest mass glueball state switches 
to a higher mass state beyond the small $\beta$ minimum. Although we do
not have an explanation for this, presumably, by including
additional states in the minimisation basis or implementing a more
complicated vacuum state, the level crossing would no longer appear
and the small $\beta$ minima would extend into large $\beta$ scaling regions. This should be checked
in further studies. A glueball mass
extracted from a large $\beta$
scaling region can be confidently compared to a corresponding
Lagrangian calculation, for in that region of couplings the ratio of
$g_E^2$ to $g^2$ is unity. 

\begin{figure}
\centering

\includegraphics[width=10cm]{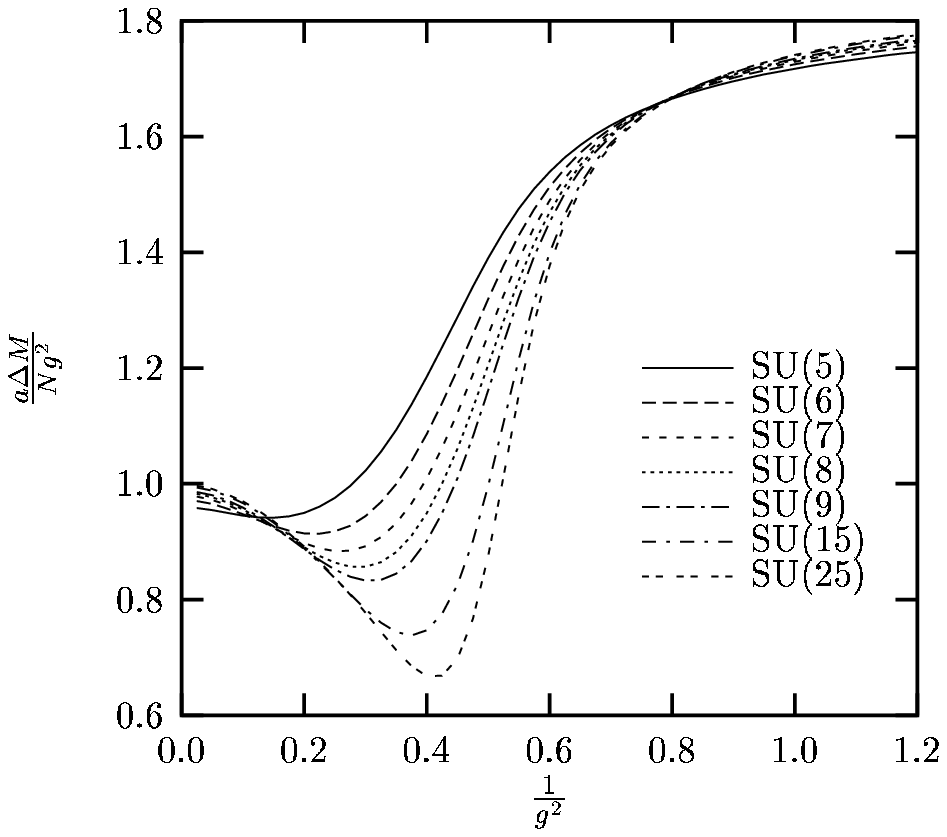}

\caption{$L_{max} = 6$ 
SU($N$) lowest mass anti-symmetric massgaps in 2+1 dimensions in units of
$g^2 N/a$ as functions of $1/g^2$. The $l$-sum of
\eqns{coolsum}{coolersum} truncated at $l_{{\rm max}}=5$. } 
\label{smallbetamin-asym}  
\end{figure}

\begin{figure}
\centering
                       
\includegraphics[width=10cm]{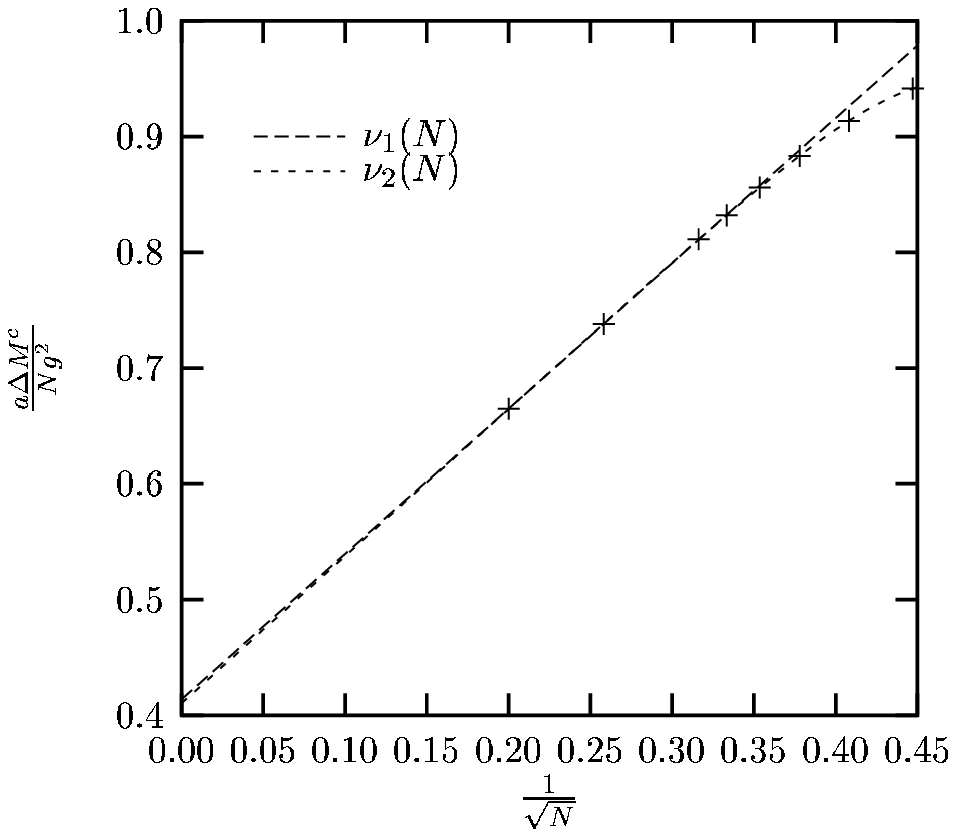}

\caption{ The continuum limit lowest mass antisymmetric 2+1
dimensional massgaps in units of $N g^2/a$ as a
function of $1/\sqrt{N}$ taken from the small $\beta$ minima. The dashed lines are fits to \eqns{smallbetafit1-asym}{smallbetafit2-asym}.}
\label{smallbetalargen-asymmetric}  
\end{figure}

We obtain small $\beta$ minima for the $0^{--}$ massgaps but only
for $N\ge 5$. Plots of these massgaps with $N\ge 5$ are
shown in \fig{smallbetamin-asym}. The behaviour is
significantly different to that observed for the symmetric
massgap. By $N=25$ the minima do not appear close to
convergence. Indeed if convergence is occurring at all, it is
significantly slower than was observed for the $0^{++}$ state. 
To explore this further, in
\fig{smallbetalargen-asymmetric} 
we plot the minima of
\fig{smallbetamin-asym}, which we again denote by $a\Delta
M^c/(N g^2)$, as a  function of $1/\sqrt{N}$. We observe linear
behaviour in the large $N$ limit. Fitting the $N\ge 9$ data to the model
\bea
\frac{a \Delta M^c}{N g^2} = \gamma_1 + \frac{\gamma_2}{N^{\gamma_3 /2}}, 
\eea
gives best fit parameters
\bea
\gamma_1  &=& 0.41 \pm 0.01  \nn\\
\gamma_2  &=& 1.25 \pm 0.02      \nn\\
\gamma_3  &=& 0.98 \pm 0.04 ,
\label{fitone}
\eea
which is consistent with $\gamma_3 = 1$. Fitting the $N \ge 9$ data to a model with
leading order finite $N$ corrections starting at $1/\sqrt{N}$ gives
\bea
\nu_1(N) = 0.41390 \pm 0.00007 + \frac{1.255 \pm 0.003}{\sqrt{N}}.
\label{smallbetafit1-asym}
\eea 
As was done for the symmetric case we can attempt to go further and
find the form of the next to leading order finite $N$ corrections.
Fitting the $N\ge 5$ data to the model
\bea
\nu_2(N) = \gamma_1 + \frac{\gamma_2}{\sqrt{N}} +
\frac{\gamma_3}{N^{\gamma_4 /2}}. 
\label{smallbetafit2-asym}
\eea
gives best fit parameters 
\bea
 \gamma_1 &=&      0.410     \pm    0.001  \\
\gamma_2 &=&  1.2718      \pm    0.0045     \\
\gamma_3 &=&  -76     \pm     23     \\
\gamma_4 &=& 9.5      \pm    0.4  
\eea
indicating vanishingly small next to leading order corrections for
$N\ge 5$. The models $\nu_1$ and $\nu_2$ are plotted against the $N\ge
5$ data in \fig{smallbetalargen-asymmetric}. 

The small $\beta$ minima appear to converge to a non-zero
$N\rightarrow \infty$ limit. The convergence is significantly slower
than for the small $\beta$ minima in the symmetric case. It is also
clear that the leading order finite $N$ corrections are not the
expected $\ord(1/N^2)$ but rather $\ord(1/\sqrt{N})$. The
$N\rightarrow \infty$ limit of $0.41 \pm 0.01$ is approximately half
the corresponding $0^{++}$ result. Such a state does not appear in the
calculation of Teper. 

It is possible that the small $\beta$ minima are spurious. Dimensional
analysis gives the expected scaling form but 
does not allow us to decide which scaling region is
preferable, or if one is a lattice artifact. 
However, based on lattice calculations to date, 
 we expect the $0^{++}$ glueball to be lighter than the $0^{--}$ in the continuum limit. 
Spurious scaling regions have been observed in other lattice calculations~\cite{WichmannPhD}.
It is clear that additional analysis is needed
here. Again, an important step will be to include additional non-rectangular
states in the minimisation basis.

\subsection{The large $\bm{\beta}$ plateaux}

Having considered the small $\beta$ minima as possible scaling regions
we now move on to the large $\beta$ plateaux. This scaling  region
appears for all $N$ and for all states considered and so its
interpretation as a genuine scaling region is less
dubious. Furthermore, the glueball mass results extracted from these scaling
regions may be confidently compared to the corresponding Lagrangian
calculations since the ratio of the Hamiltonian to Euclidian coupling
is unity up to small $\ord(g_E^2)$ corrections.

\begin{figure}
\centering
                       
\includegraphics[width=10cm]{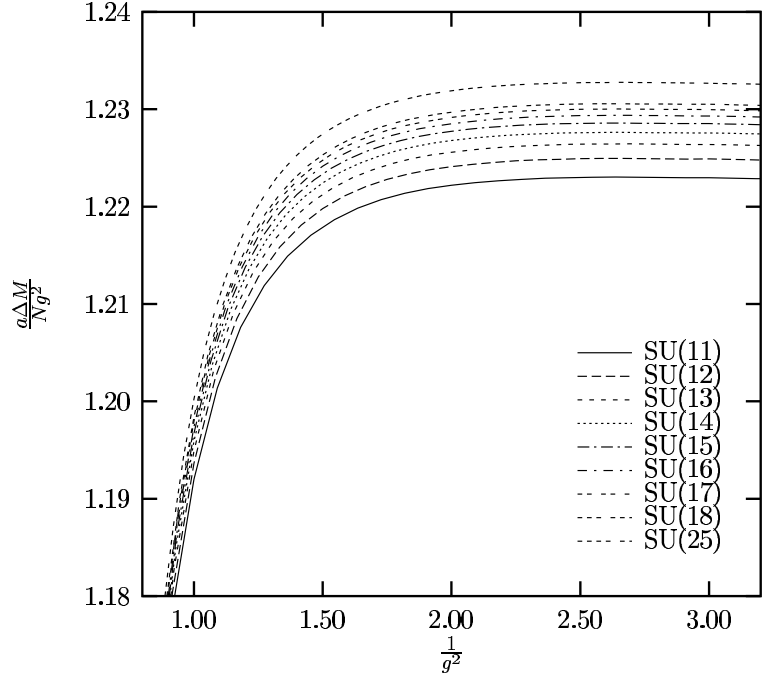}

\caption{ Lowest mass symmetric 2+1 dimensional massgaps in units
of $N g^2/a$ as a function of $1/g^2$.}
\label{S-ev1-hib-CONV}  
\end{figure}

\begin{figure}
\centering
                       
\includegraphics[width=10cm]{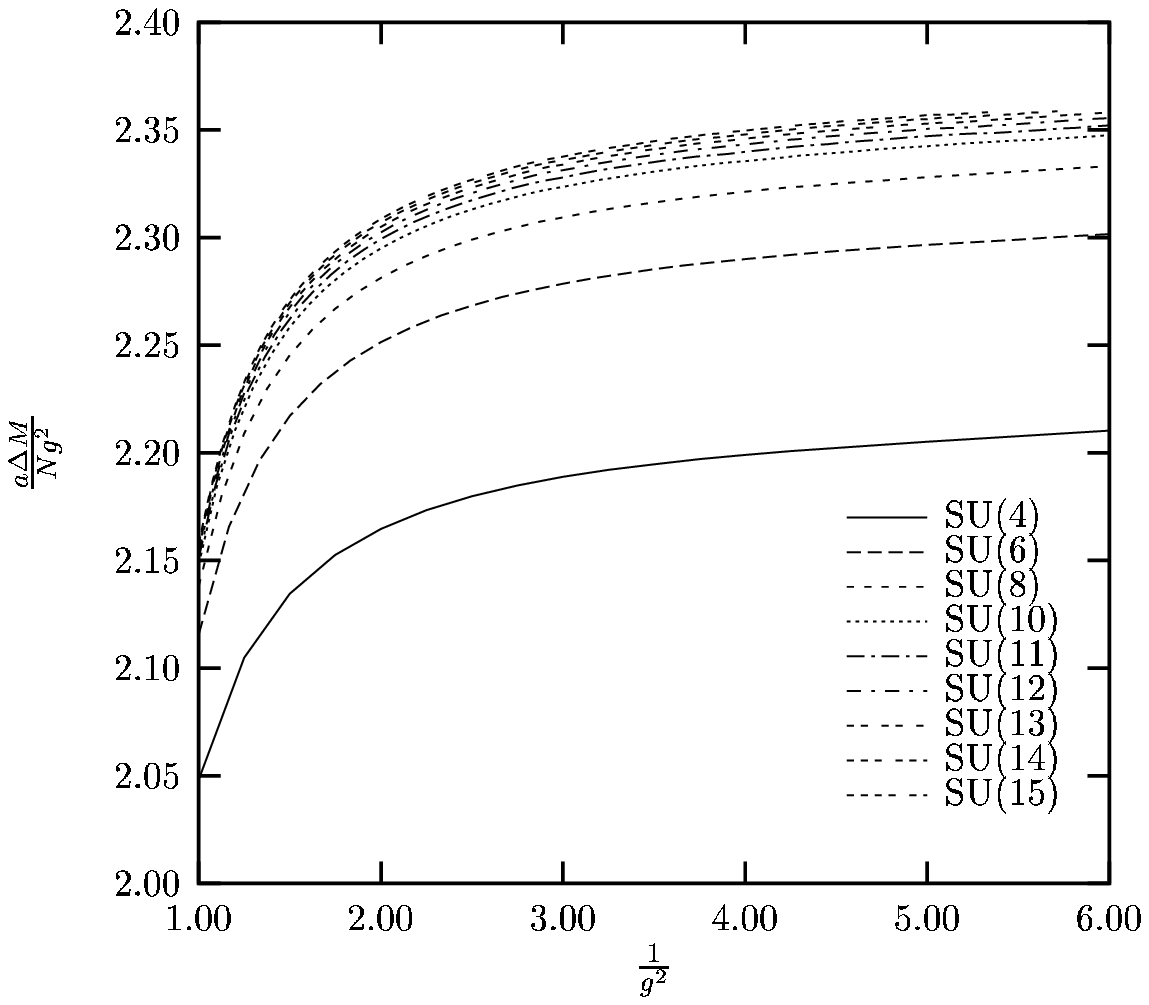}

\caption{Second lowest mass symmetric 2+1 dimensional massgaps in units
of $N g^2/a$ as a function of $1/g^2$.}
\label{S-ev2-hib-CONV}  
\end{figure}

We start with the $0^{++}$ state for which the best scaling behaviour
is obtained. We calculate the five lowest lying massgaps corresponding
to the five lowest glueball masses accessible with our choice of 
ground state and minimisation
basis. We consider values of $N$ in the range $3\le N \le 25$. For each $N$ considered 
we find a large $\beta$ scaling plateau for each of the five lowest mass states. 
The lowest lying massgap is shown in \fig{S-ev1-hib-CONV} for a range
of $N$. We observe that in units of $N g^2/a$ the massgaps do not
depend strongly on $N$ and that in the scaling region they appear to
approach a finite limit. Similar observations can be made for the
four higher mass states obtained although the scalaing behaviour is
less precise. We show the second lowest glueball mass in \fig{S-ev2-hib-CONV} as an example. A continuum limit for each massgap
is obtained in the scaling region by fitting to a constant. For each
fit we use a region of at least 10 data points which minimises the
standard error. The continuum limit results for the five lowest lying
massgaps, denoted by  $a
\Delta M^c /(N g^2)$, are shown as functions of $1/N^2$ in
\figs{S-ev1-hib-CONV}{s-largen-conv}. Also shown in
the plots are fits to models with leading order large $N$ corrections
starting $1/N^2$,
\bea
\kappa_i^{++} = p_i + \frac{q_i}{N^2} +  \frac{r_i}{N^4}.
\label{kappa++}
\eea 
Here the subscript $i$ labels the $i$-th lowest glueball mass. 
The values of the best fit parameters are given in \tab{kappa++fit}.

\begin{table}
\caption{The best fit parameters for the five lowest energy $0^{++}$ massgaps
when fitting \eqn{kappa++} to the available data.}
\label{kappa++fit}
\begin{ruledtabular}
\begin{tabular}{cD{/}{\mbox{$\pm$}}{-2}D{/}{\mbox{$\pm$}}{-2}D{/}{\mbox{$\pm$}}{-2}}
\multicolumn{1}{c}{$i$}  &  \multicolumn{1}{c}{$p_i$}   &  \multicolumn{1}{c}{$q_i$}  & \multicolumn{1}{c}{$r_i$}   \\
\hline
1  &  1.23526/ 0.00023    & -1.540/ 0.018      &  1.97 /  0.16    \\
2 & 2.36924 / 0.00072      & -2.478/ 0.047      & -1.62 / 0.42       \\
3 &  2.88446 / 0.00071      & -3.435 / 0.017      &0\footnotemark[1]       \\
4 &   3.35422 / 0.00047     &  -3.476 / 0.012     & 0\footnotemark[1]      \\
5 &   3.7667 / 0.0013     &  -4.114 / 0.092     & 1.88 /
0.83 \\
\end{tabular}
\end{ruledtabular}
\footnotetext[1]{Set to zero to obtain a stable fit.}
\end{table} 

\begin{figure}
\centering
                       
\includegraphics[width=10cm]{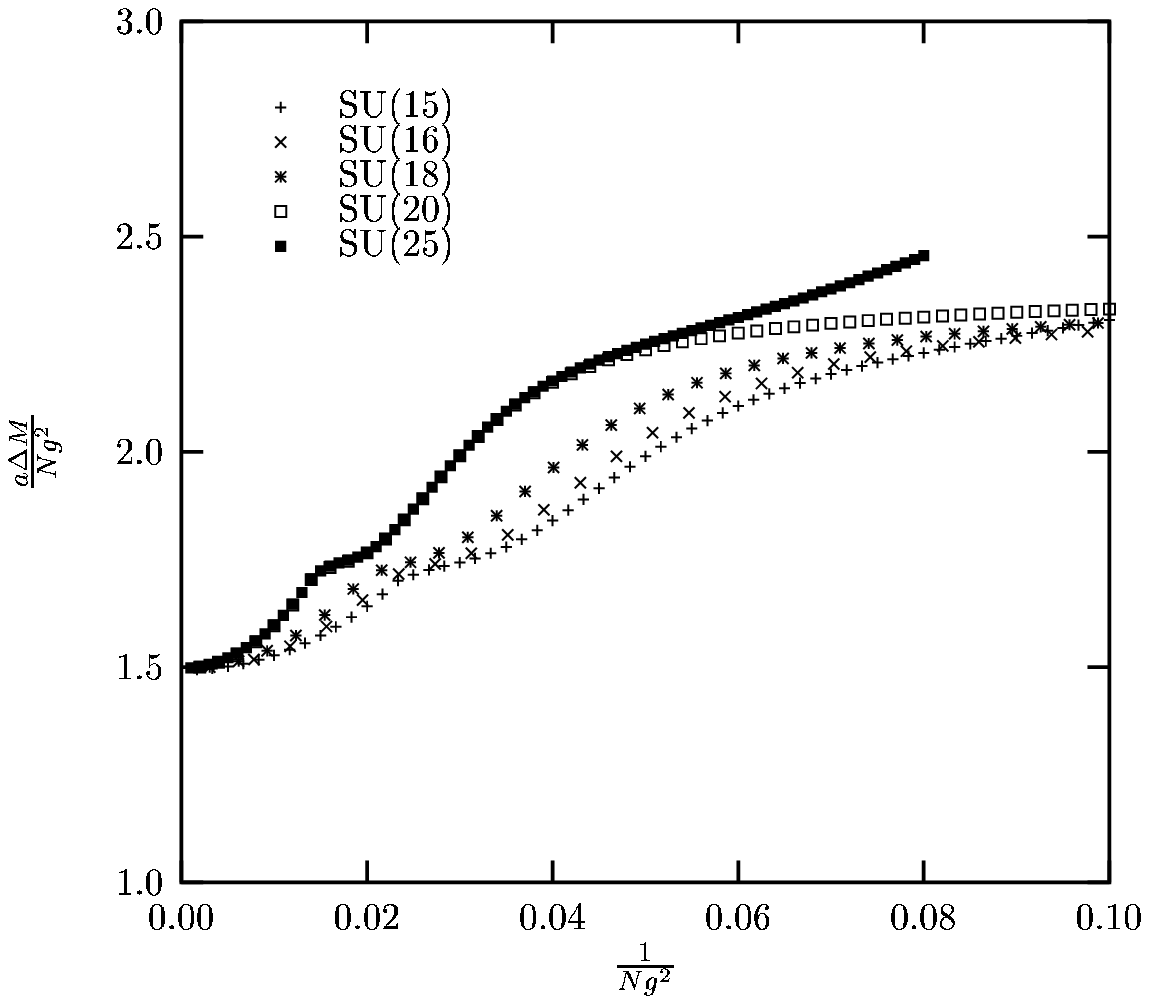}

\caption{Second lowest mass symmetric 2+1 dimensional massgaps in units
of $N g^2/a$ as a function of $1/(N g^2)$.}
\label{S-ev2-lob-CONV}  
\end{figure}

These results should be compared with those of Lucini and Teper who
obtain large $N$ glueball masses in units of $N g^2/a$ in the $0^{++}$ sector with linear fits given by
\bea
 0^{++}:&& 0.8116(36) - \frac{0.090(28)}{N^2} \nn\\
 0^{++*}:&& 1.227(9) - \frac{0.343(82)}{N^2} \nn\\
 0^{++**}:&& 1.65(4) -\frac{2.2(7)}{N^2}.
\eea
The fit for the $0^{++**}$ state is obtained using $4\le N \le 6$
data. The remaining fits are obtained using $2\le N\le 6$ data.
We find that the lowest glueball mass extracted from our large $\beta$ 
plateaux 
is consistent, in the $N\rightarrow \infty$ limit, with the
state which Lucini and Teper label $0^{++**}$. The slopes of the fits are of the same
sign but differ significantly in magnitude. The $0^{++**}$ state of
Lucini and Teper does not appear in our data in the form of a large
$\beta$ scaling plateau. There is however a hint of an approach to
scaling in the vicinity of their result in our second lowest massgap data. This
effect is only visible in our data for $N \ge 13$ and occurs for small $\beta$
as shown in \fig{S-ev2-lob-CONV}. The values of $\beta$ for
which this effect appears are quite close to those for which the small
beta minima are observed in the lowest mass eigenstate. 
Similar effects are observed
in the $0^{--}$ data but the effect is much less convincing with the
data currently available.

\begin{figure}
\centering
                       
\includegraphics[width=10cm]{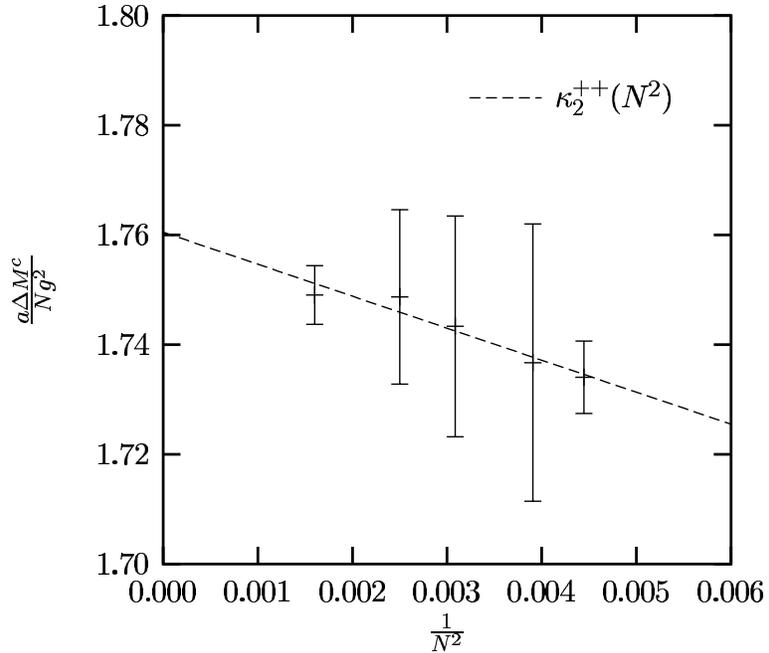}

\caption{Continuum limit extrapolations (in units
of $N g^2/a$ as a function of $1/N^2$) derived from the low $\beta$ scaling
region that appears for $N\ge 13$ in the second lowest mass $0^{++}$
SU($N$) massgap. The dashed line is the fit to the
quadratic model of \eqn{kappa++lob}.}
\label{S-ev2-lob}  
\end{figure}

Let us consider this small $\beta$ region as a possible scaling
region. We fit a constant to the available $N\ge 15$ data on a range
of at least 4 data points chosen so that the standard error is
minimised. These scaling values are plotted as a function of $1/N^2$
in \fig{S-ev2-lob}. Also shown is the best fit linear model
\bea
\kappa_2^{++} = 1.7605\pm 0.0032 - \frac{5.83 \pm 0.97}{N^2}. 
\label{kappa++lob}
\eea
This produces an $N\rightarrow \infty$ limit which is close to but
inconsistent with the result obtained by Lucini and Teper for their 
$0^{++**}$ state.

\begin{figure}
\centering
                       
\includegraphics[width=10cm]{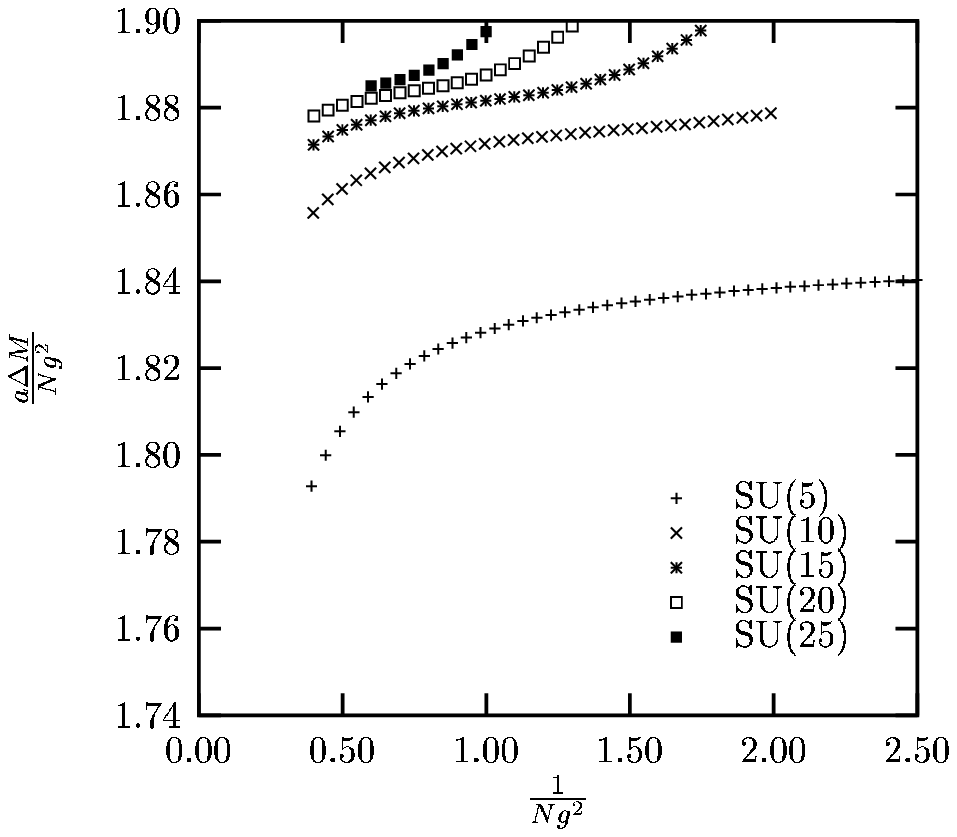}

\caption{ Lowest mass $0^{--}$ 2+1 dimensional massgaps in units
of $N g^2/a$ as a function of $1/(N g^2)$.}
\label{AS-ev1-hib-CONV}  
\end{figure}

\begin{figure}
\centering
                       
\includegraphics[width=10cm]{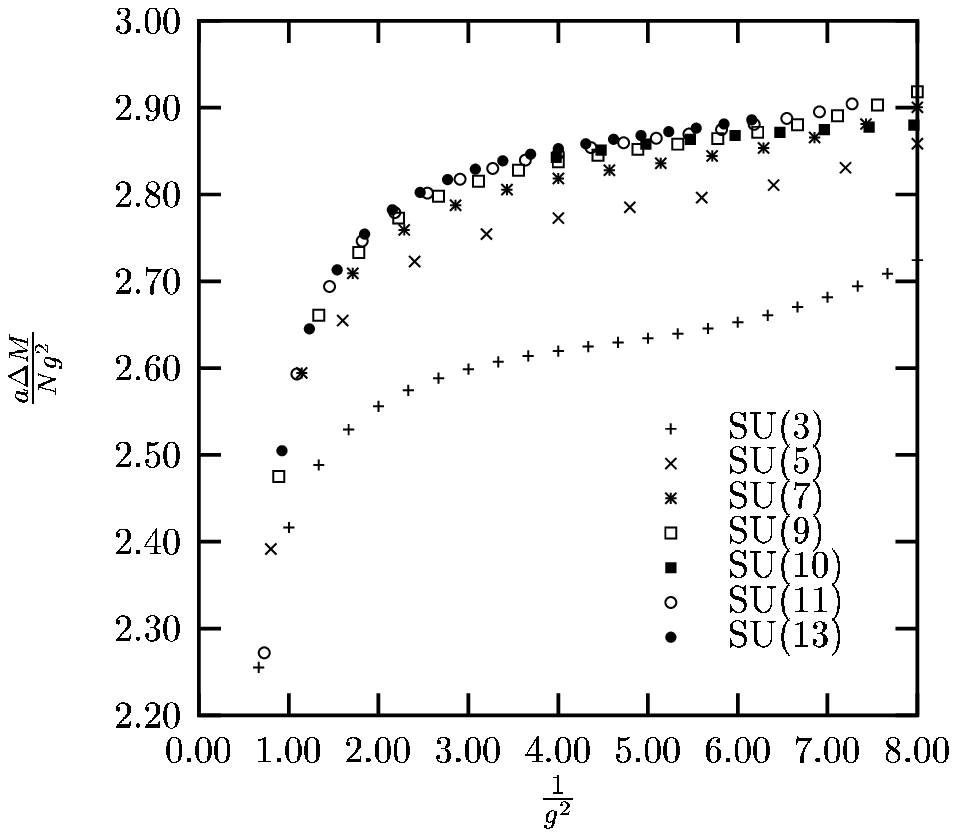}

\caption{Second lowest mass $0^{--}$ 2+1 dimensional massgaps in units
of $N g^2/a$ as a function of $1/g^2$.}
\label{AS-ev2-hib-CONV}  
\end{figure}

\begin{figure}
\centering
                       
\includegraphics[width=10cm]{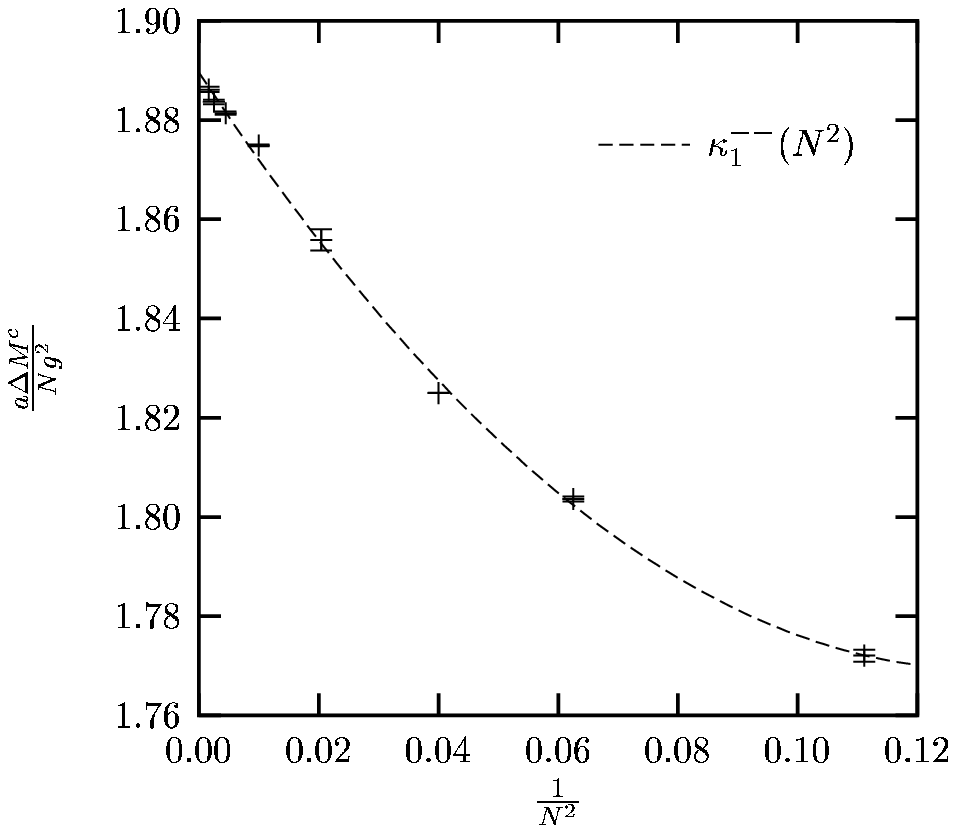}

\caption{The 2+1 dimensional continuum limit lowest $0^{--}$ SU($N$)
glueball mass
 in units
of $N g^2/a$ as a function of $1/N^2$. The dashed line is the fit to the
quadratic model of \eqn{kappa--}.}
\label{AS-ev1-hib}  
\end{figure}

We now move on to the $0^{--}$ states. For these states the scaling
behaviour is not as precise as that obtained for the $0^{++}$ states.  
Again we calculate the five lowest lying massgaps corresponding
to the five lowest mass glueballs accessible with our choice of minimisation
basis and ground state. For each $N$ considered up to 25 we find a
large $\beta$ scaling plateau for each of the five states. The lowest
lying massgap is shown in \fig{AS-ev1-hib-CONV} for a range
of $N$. The second lowest energy massgap is shown in
\fig{AS-ev2-hib-CONV}. Considerably less data has been obtained
for the $0^{--}$ states due to the large $l_{\rm max}$ required for
convergence as discussed in \sect{convwithl}. 
Despite this, the results qualitatively replicate those
of the $0^{++}$ states. We observe that in units of $N g^2/a$ the
massgaps do not depend strongly on $N$ and 
that in the scaling region they appear to
approach a finite limit. Similar observations can be made for the
four higher mass states although the scaling behaviour is less precise. The continuum limit values extracted
are plotted as functions of $1/N^2$ in 
\figs{AS-ev1-hib}{as-largen-conv}. The dashed lines show fits of the
continuum limit values to the model 
\bea
\kappa_i^{--} = p_i + \frac{q_i}{N^2} +  \frac{r_i}{N^4},
\label{kappa--}
\eea
with the parameters for each excited state given in
\tab{kappa--fit}.

\begin{table}
\caption{The best fit parameters for the five lowest energy $0^{--}$ massgaps
when fitting \eqn{kappa--} to the available data.}
\label{kappa--fit}
\begin{ruledtabular}
\begin{tabular}{cD{/}{\mbox{$\pm$}}{-2}D{/}{\mbox{$\pm$}}{-2}D{/}{\mbox{$\pm$}}{-2}}
\multicolumn{1}{c}{$i$}  &  \multicolumn{1}{c}{$p_i$}   &  \multicolumn{1}{c}{$q_i$}  & \multicolumn{1}{c}{$r_i$}   \\
\hline
1  &  1.8896 / 0.0011    & -1.829 / 0.068      &
6.95 / 0.62    \\
2  &  2.8930 / 0.0044     & -2.96 / 0.23      &
6.6/ 2.0    \\
3  &  3.20871 / 0.0047    & -3.35 / 0.22      &
8.8 / 1.9    \\
4  &  3.83647 / 0.0024     & -3.41/ 0.05      &  0\footnotemark[1]   \\
5  &  4.0759 / 0.0019     & -2.45 / 0.12      &
-7.4/ 1.1    \\
\end{tabular}
\end{ruledtabular}
\footnotetext[1]{Set to zero to obtain a stable fit.}
\end{table} 

These results should be compared again with those of Lucini and Teper who
obtain large $N$ glueball masses in units of $N g^2/a$ in the $0^{--}$ sector with linear fits given by
\bea
 0^{--}:&& 1.176(14) + \frac{0.14(20)}{N^2} \nn\\
 0^{--*}:&& 1.535(28) - \frac{0.35(35)}{N^2} \nn\\
 0^{--**}:&& 1.77(13) + \frac{0.24(161)}{N^2}.
\eea
The result for $0^{--**}$ is from \rcite{Teper:1998te}. All
fits were obtained using $3\le N \le 6$ data. The $N\rightarrow
\infty$ limit of our lowest lying massgap
extracted from the large $\beta$ plateaux is consistent with Teper's
$0^{--**}$ result.

Having considered spin 0 states we now move on to spin 2, the only
other spin accessible when using a basis of rectangular states. For
the case of spin 2 the scaling behaviour of the massgaps is
significantly worse than for spin 0. More troublesome is the fact that 
the convergence of the $2^{++}$
massgaps with increasing $l_{\rm max}$ is slower than the case of $0^{--}$. 
The situation for the $2^{--}$ state is markedly better with the
convergence of massgaps with increasing $l_{\rm max}$ being no different to that of
$0^{++}$. For this reason we concentrate on the $2^{--}$ sector here.

\begin{figure}
\centering
                       
\includegraphics[width=10cm]{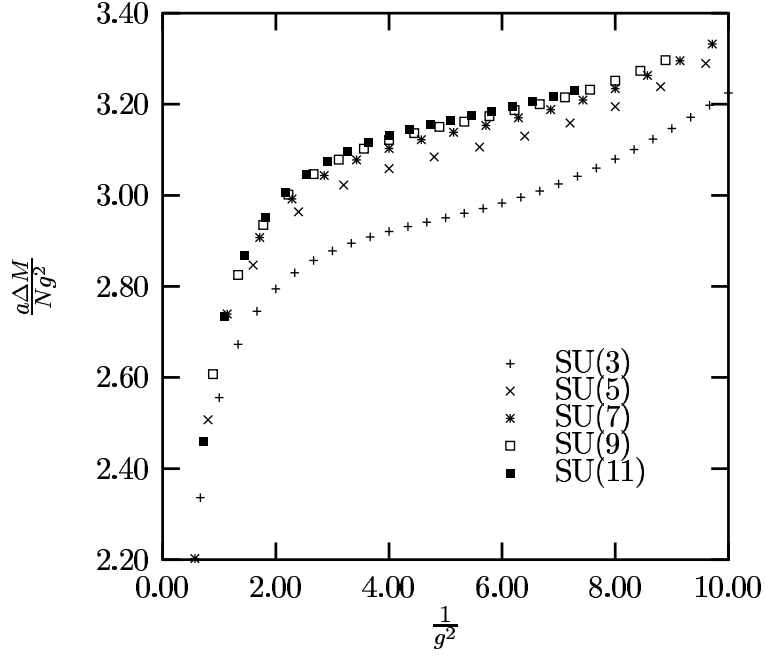}

\caption{The lowest lying 2+1 dimensional $2^{--}$ SU($N$) massgaps in units
of $N g^2/a$ as a function of $1/g^2$.}
\label{AS-ev1-hib-spin2-CONV}  
\end{figure}

\begin{figure}
\centering
                       
\includegraphics[width=10cm]{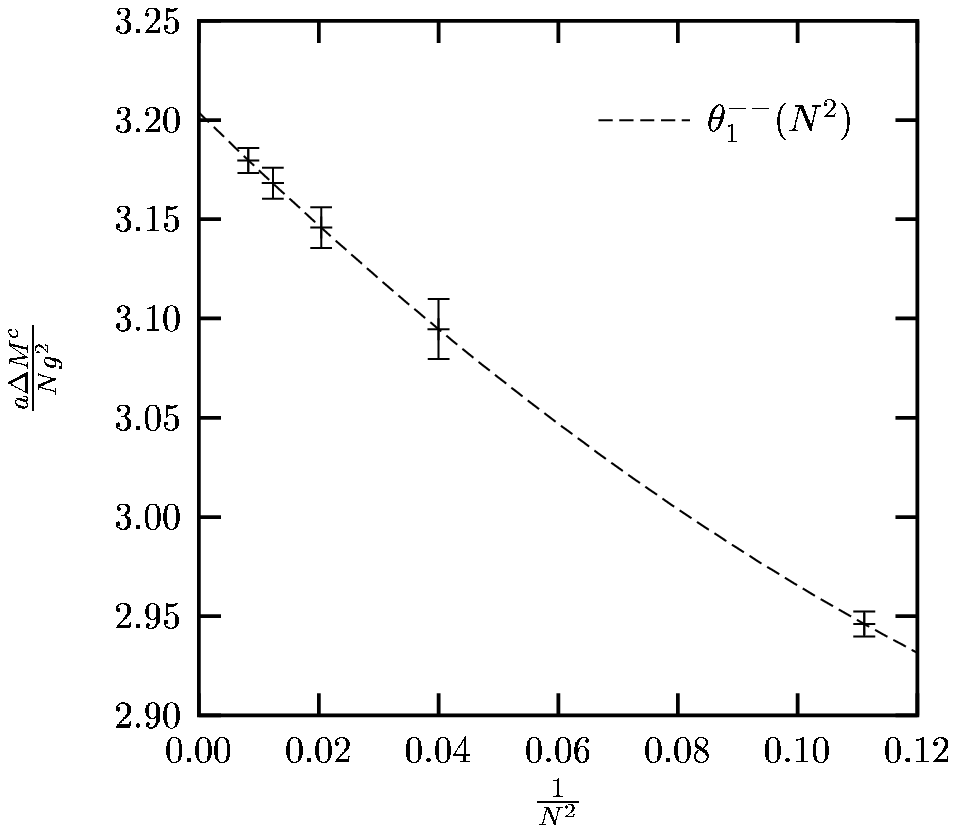}

\caption{The 2+1 dimesnional 
continuum limit lowest $2^{--}$ SU($N$) glueball mass
 in units
of $N g^2/a$ as a function of $1/N^2$. The dashed line is the fit to the
quadratic model of \eqn{theta--}.}
\label{AS-ev1-hib-spin2}  
\end{figure}

The $2^{--}$ states produce less precise scaling
behaviour than the spin 0 states. However large $\beta$ plateaux
appear for each of the five lowest lying massgaps. We use these
regions to estimate their respective continuum limits. For these states
small $\beta$ minima do not appear.  
The lowest lying $2^{--}$ massgap is shown in
\fig{AS-ev1-hib-spin2-CONV} for a range of $N$.  
Once again we observe that in units of $N g^2/a$ the massgaps do not
depend strongly on $N$ and that in the scaling region they appear to
approach a finite limit. Similar observations can be made for the
four higher mass states obtained, although the scaling behaviour
worsens as the mass of the state increases. The estimated continuum limit
values 
extracted are plotted as functions of $1/N^2$ in \figs{AS-ev1-hib-spin2}{as-largen-conv-spin2}. The dashed lines show fits to the model 
\bea
\theta_i^{--} = p_i + \frac{q_i}{N^2} +  \frac{r_i}{N^4}.
\label{theta--}
\eea
The best fit parameters for each excited state are given in \tab{theta--fit}. 

\begin{table}
\caption{The best fit parameters for the five lowest energy $2^{--}$ massgaps
when fitting \eqn{theta--} to the available data.}
\label{theta--fit}
\begin{ruledtabular}
\begin{tabular}{cD{/}{\mbox{$\pm$}}{-2}D{/}{\mbox{$\pm$}}{-2}D{/}{\mbox{$\pm$}}{-2}}
\multicolumn{1}{c}{$i$}  &  \multicolumn{1}{c}{$p_i$}   &  \multicolumn{1}{c}{$q_i$}  & \multicolumn{1}{c}{$r_i$}   \\
\hline
1  &  3.20379/ 0.00006    & -2.9571/ 0.0032      &
5.73912/ 0.02585    \\
2  &  4.07376/ 0.00066    & -2.189/ 0.050      &
-9.46 / 0.43    \\
3  &  4.96259 / 0.00090    & -4.947 / 0.045      &
1.99/ 0.35    \\
4  &  5.26717/ 0.00040    & -6.221/ 0.031      & 15.92
/ 0.26    \\
5  &  5.744 / 0.024    & -5.13/ 0.56       & 0\footnotemark[1]   \\
\end{tabular}
\end{ruledtabular}
\footnotetext[1]{Set to zero to obtain a stable fit.}
\end{table} 

Once again these results should be compared with those of Lucini and Teper who
obtain large $N$ glueball masses in units of $N g^2/a$ in the $2^{--}$ sector with linear fits derived from
$3\le N \le 6$ data given by
\bea
 2^{--}:&& 1.615(33) - \frac{0.10(42)}{N^2} \nn\\
 2^{--*}:&& 1.87(12) - \frac{0.37(200)}{N^2}. 
\eea
The $2^{--*}$ result is from \rcite{Teper:1998te}. The 
$N\rightarrow \infty$ results presented here are significantly higher. This time 
no correspondence between our results and those of Lucini and Teper can
be obtained.  

\begin{figure}
\centering
                       
\includegraphics[width=10cm]{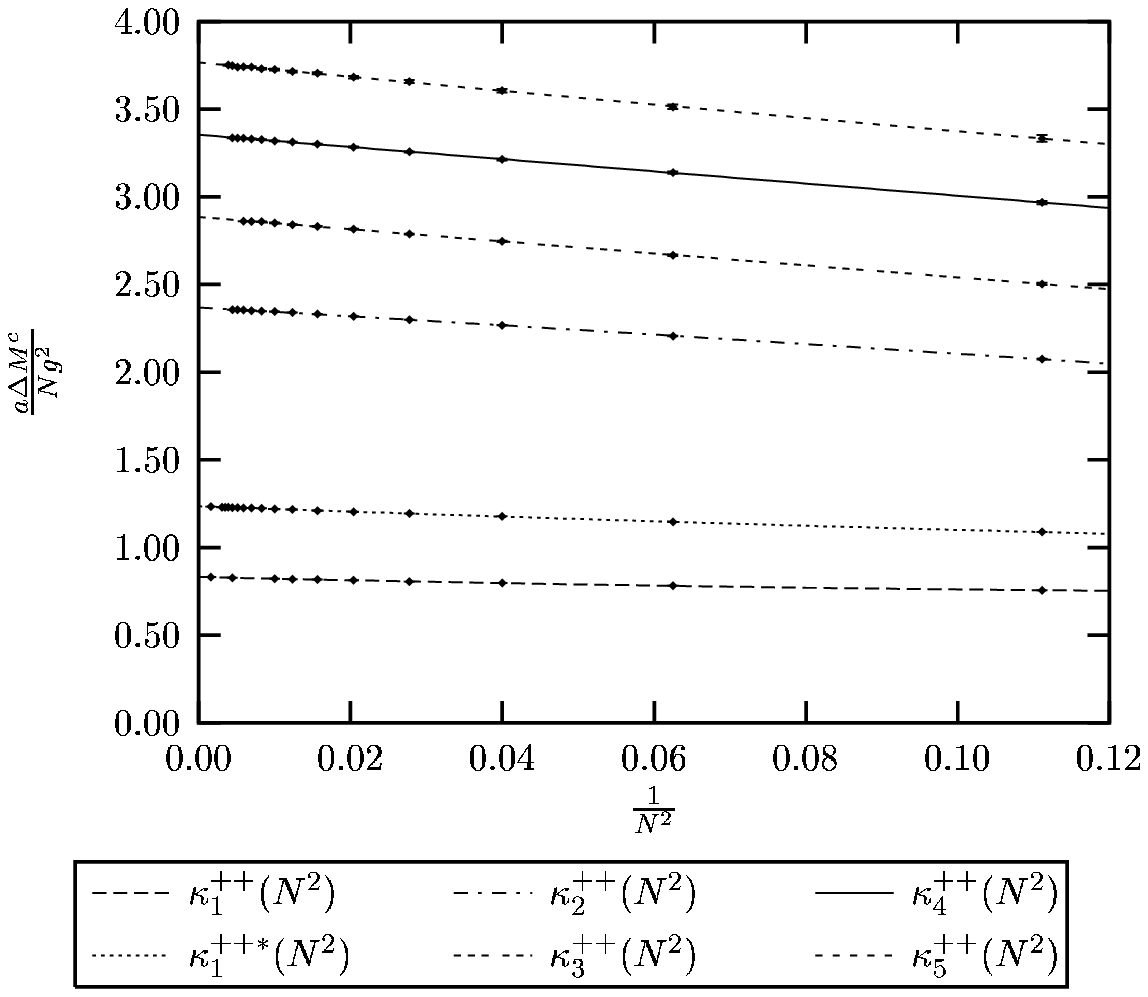}

\caption{An estimate of the continuum limit mass spectrum for 
$0^{++}$ SU($N$) glueballs in units
of $N g^2/a$ as a function of $1/N^2$. The lines are fits to the 
models of \eqn{kappa++}.}
\label{S-spectrum}  
\end{figure}

We finish this section by presenting the mass spectra obtained for the
$0^{++}$, $0^{--}$ and $2^{--}$ sectors. The results are summarised by
the plots in \figss{S-spectrum}{AS-spin2-spectrum}.

\begin{figure}
\centering
                       
\includegraphics[width=10cm]{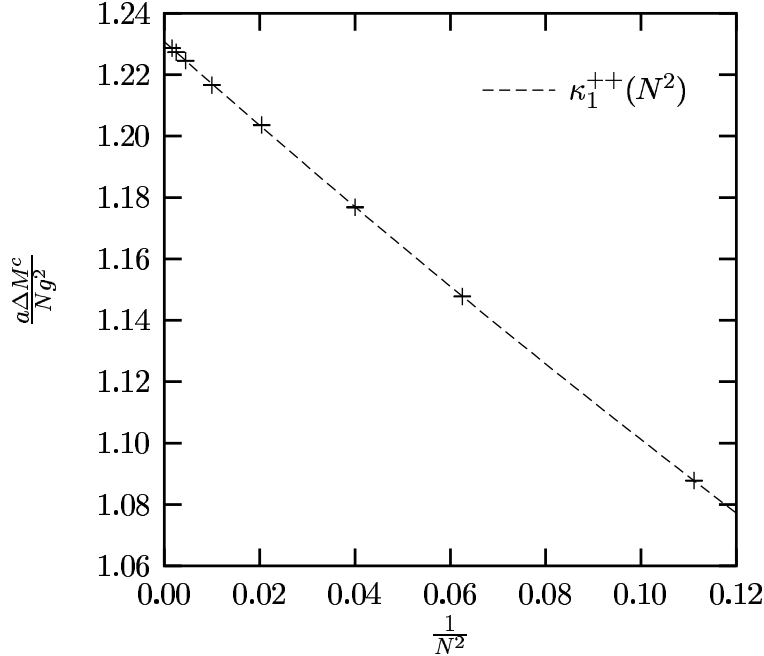}

\caption{Continuum limit lowest mass $0^{++}$ SU($N$) massgap
 in units
of $N g^2/a$ as a function of $1/N^2$. The dashed line is the fit to the
quadratic model of \eqn{kappa++}.}
\label{S-ev1-hib}  
\end{figure}

\begin{figure}
\centering
                             
\subfigure[2nd eigenvalue] 
                     {
                         \label{S-ev2-hib}
                         \includegraphics[width=7cm]{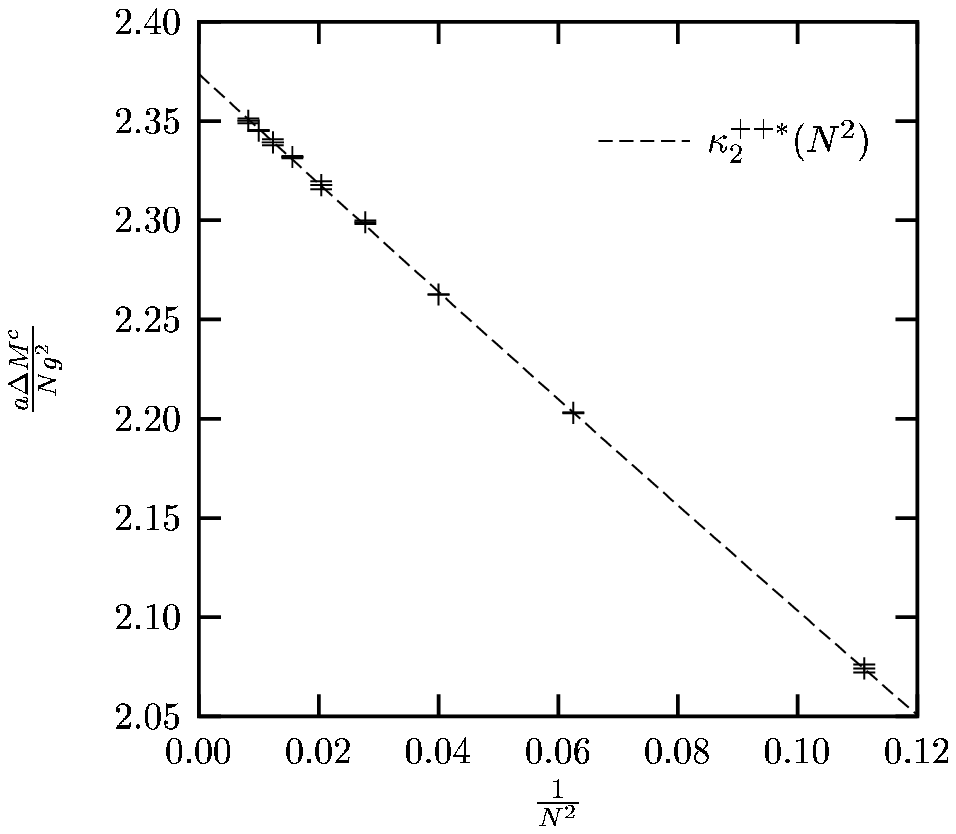}
                     }\hspace{0.25cm}
\subfigure[3rd eigenvalue] 
                     {
                         \label{S-ev3-hib}
                         \includegraphics[width=7cm]{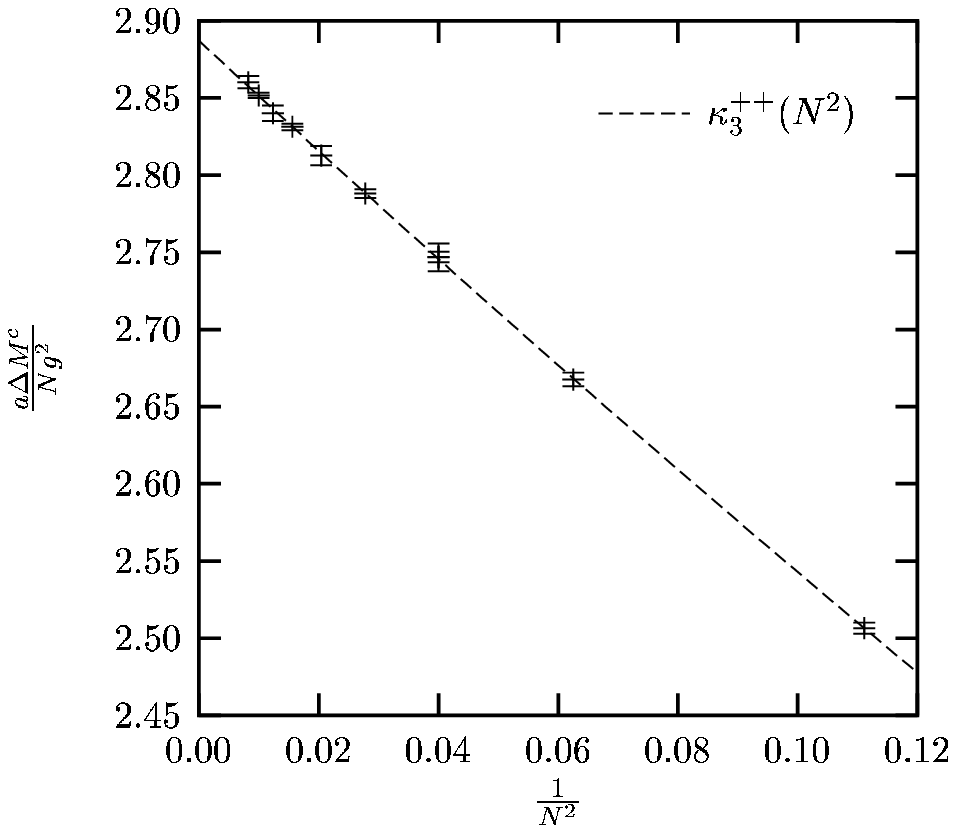}
                     }\\ 
\subfigure[4th eigenvalue] 
                     {
                         \label{S-ev4-hib}
                         \includegraphics[width=7cm]{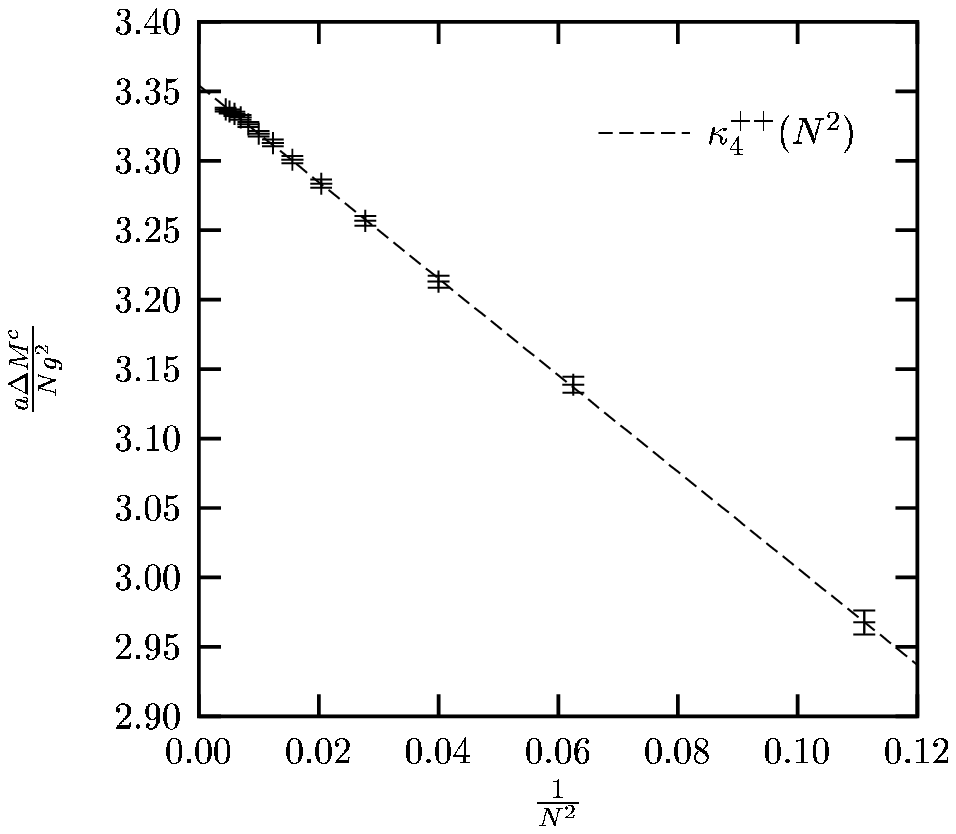}
                     }\hspace{0.25cm}
\subfigure[5th eigenvalue] 
                     {
                         \label{S-ev5-hib}
                         \includegraphics[width=7cm]{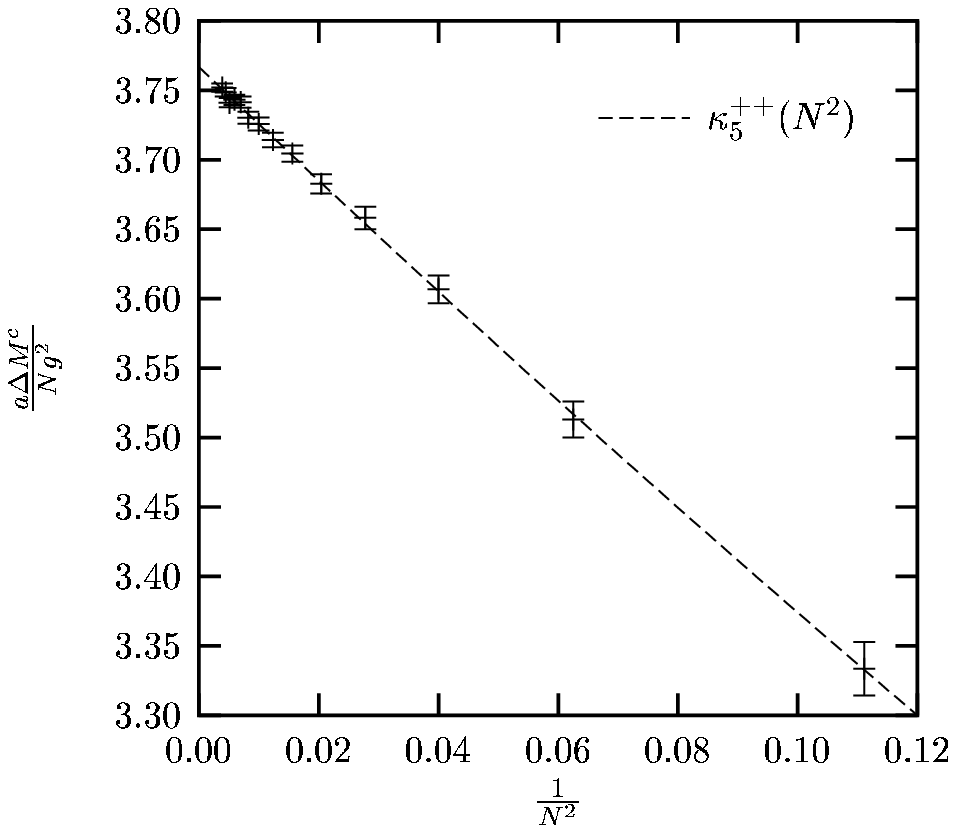}
                     }
\caption{Continuum limit $0^{++}$ SU($N$) massgaps in units of $N
 g^2/a$ as functions of $1/N^2$. The dashed lines are fits given in
\eqn{kappa++}.} 
\label{s-largen-conv}  
\end{figure}

\begin{figure}
\centering
               
\subfigure[2nd eigenvalue] 
                     {
                         \includegraphics[width=7cm]{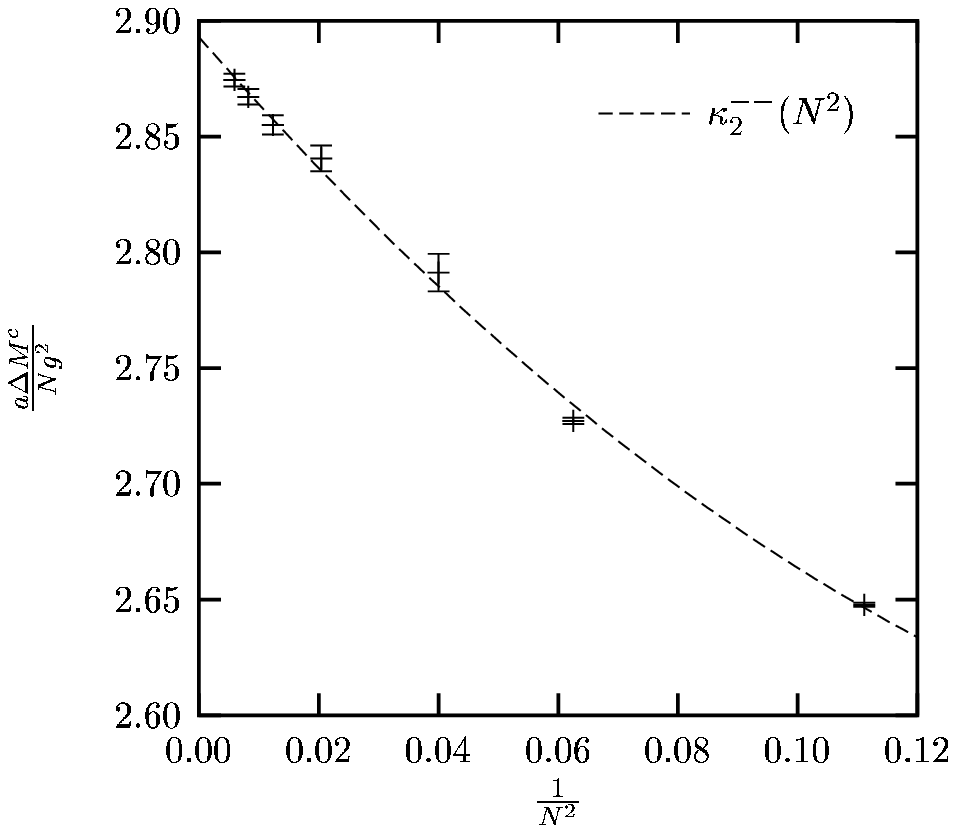}
                     }\hspace{0.25cm}
\subfigure[3rd eigenvalue] 
                     {
                         \includegraphics[width=7cm]{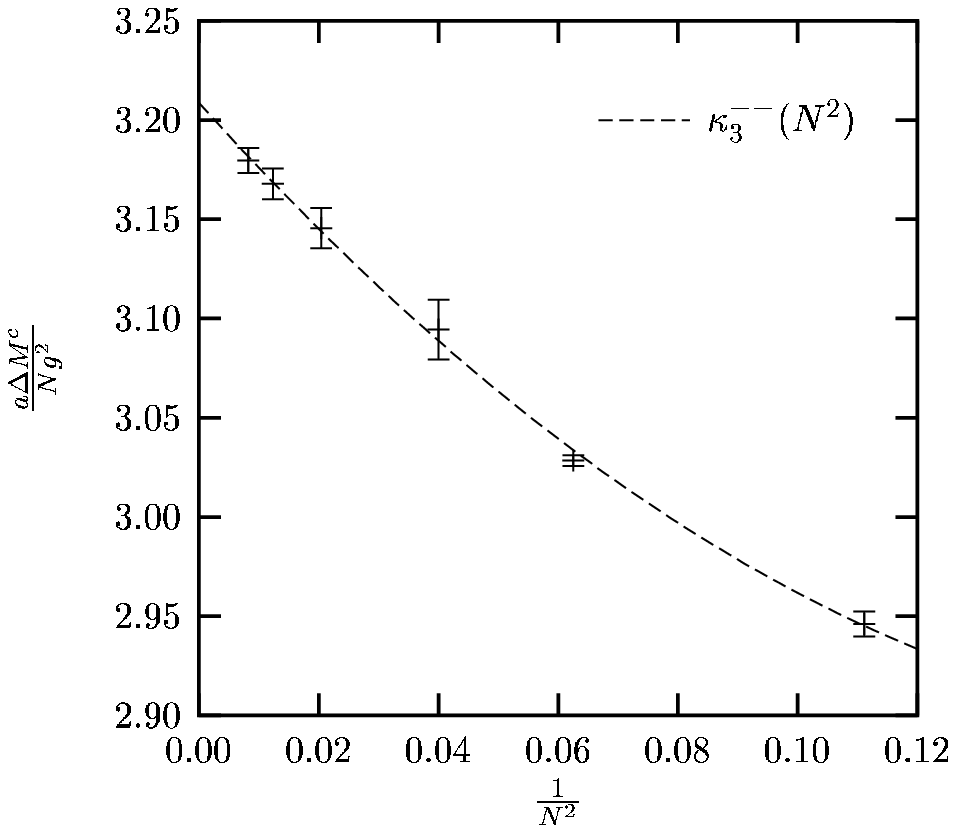}
                     }\\
\subfigure[4th eigenvalue] 
                     {
                         \includegraphics[width=7cm]{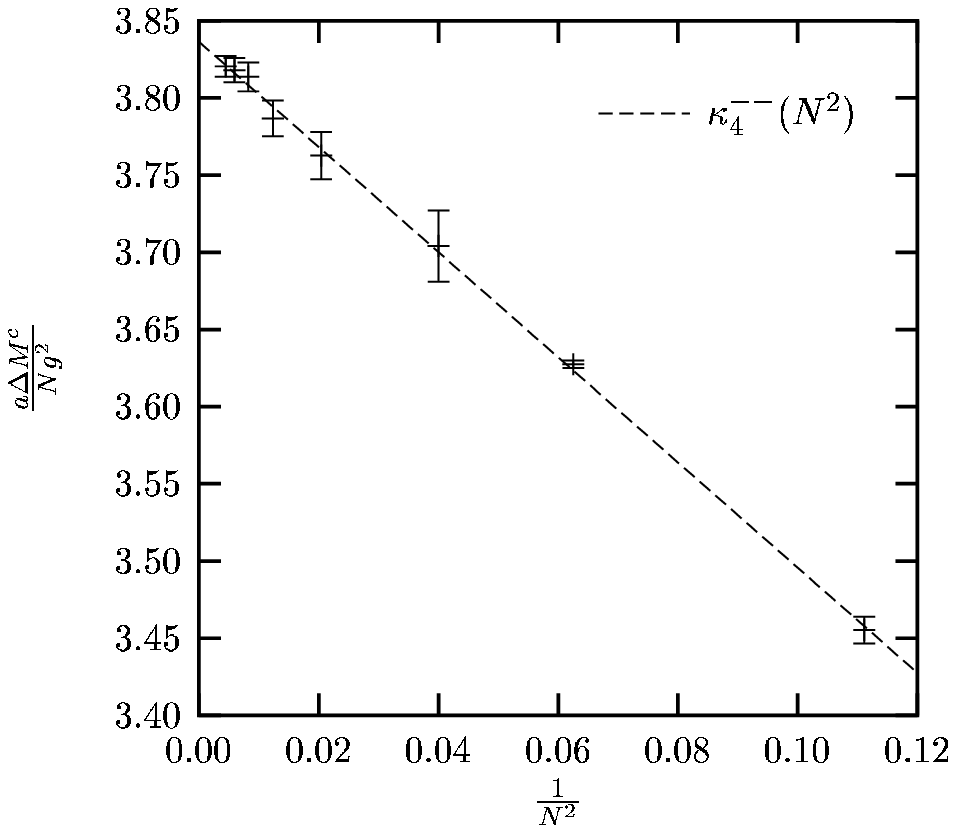}
                     }\hspace{0.25cm}      
\subfigure[5th eigenvalue] 
                     {
                         \includegraphics[width=7cm]{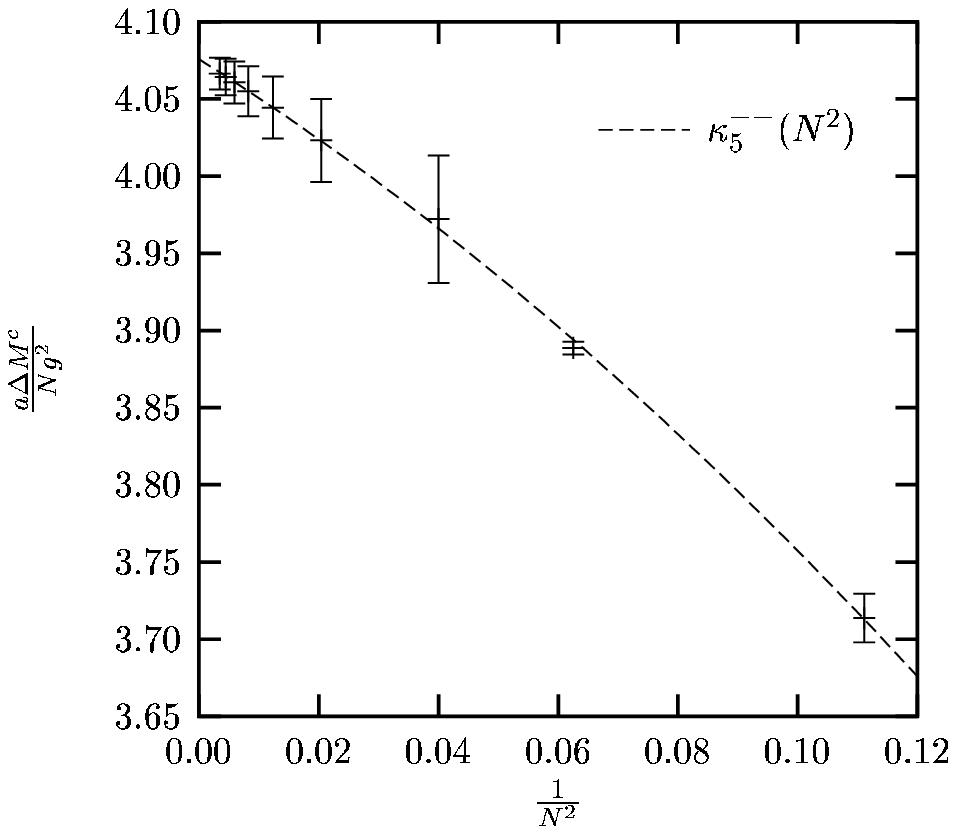}
                     }
\caption{Continuum limit $0^{--}$ SU($N$) massgaps in units of $N
 g^2/a$ as functions of $1/N^2$. The dashed lines are fits given in
 \eqn{kappa--}.} 
\label{as-largen-conv}  
\end{figure}

\begin{figure}
\centering
               
\subfigure[2nd eigenvalue] 
                     {
                         \includegraphics[width=7cm]{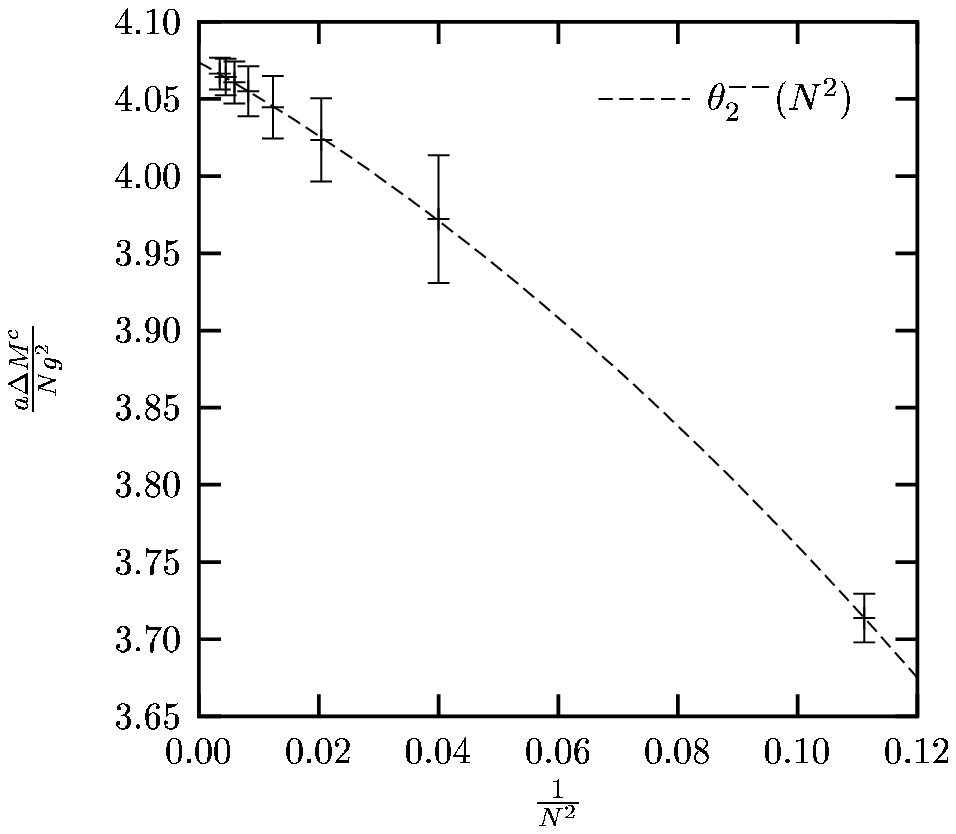}
                     }\hspace{0.25cm}
\subfigure[3rd eigenvalue] 
                     {
                         \includegraphics[width=7cm]{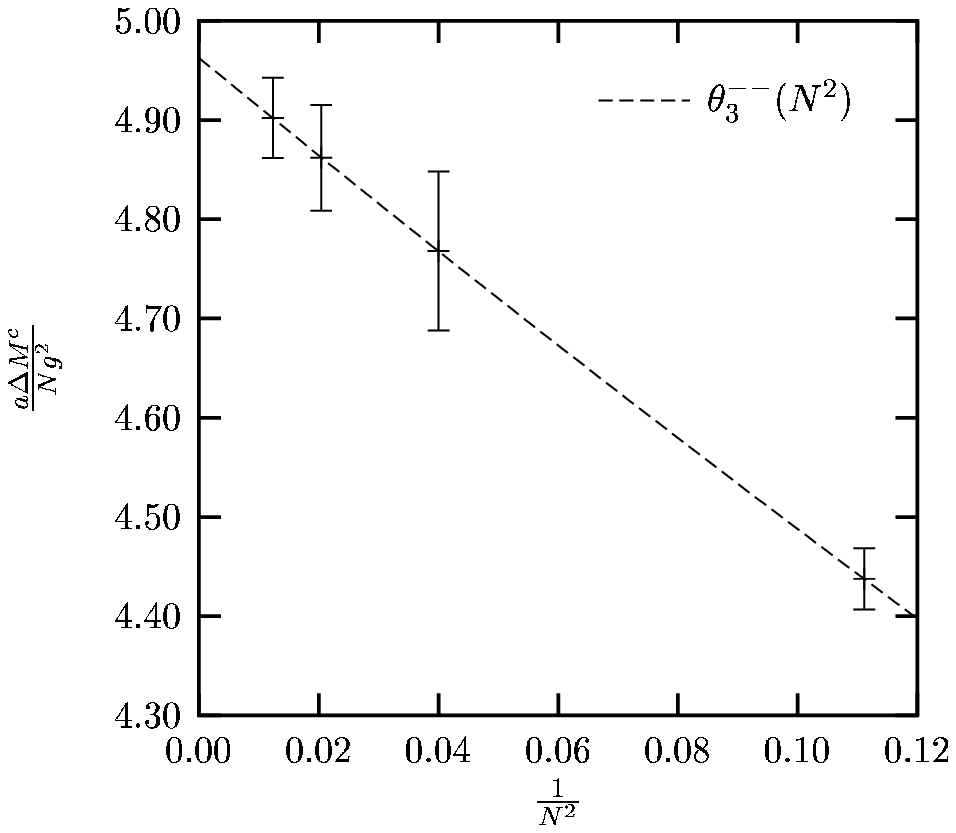}
                     }\\
\subfigure[4th eigenvalue] 
                     {
                         \includegraphics[width=7cm]{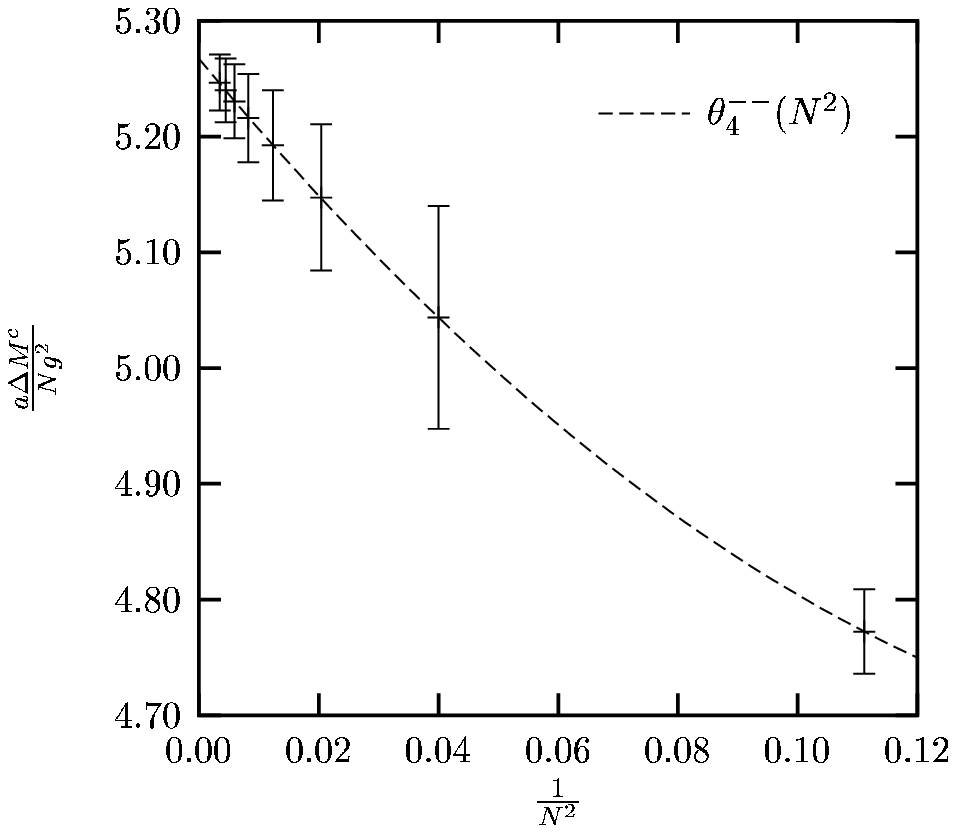}
                     }\hspace{0.25cm}      
\subfigure[5th eigenvalue] 
                     {
                         \includegraphics[width=7cm]{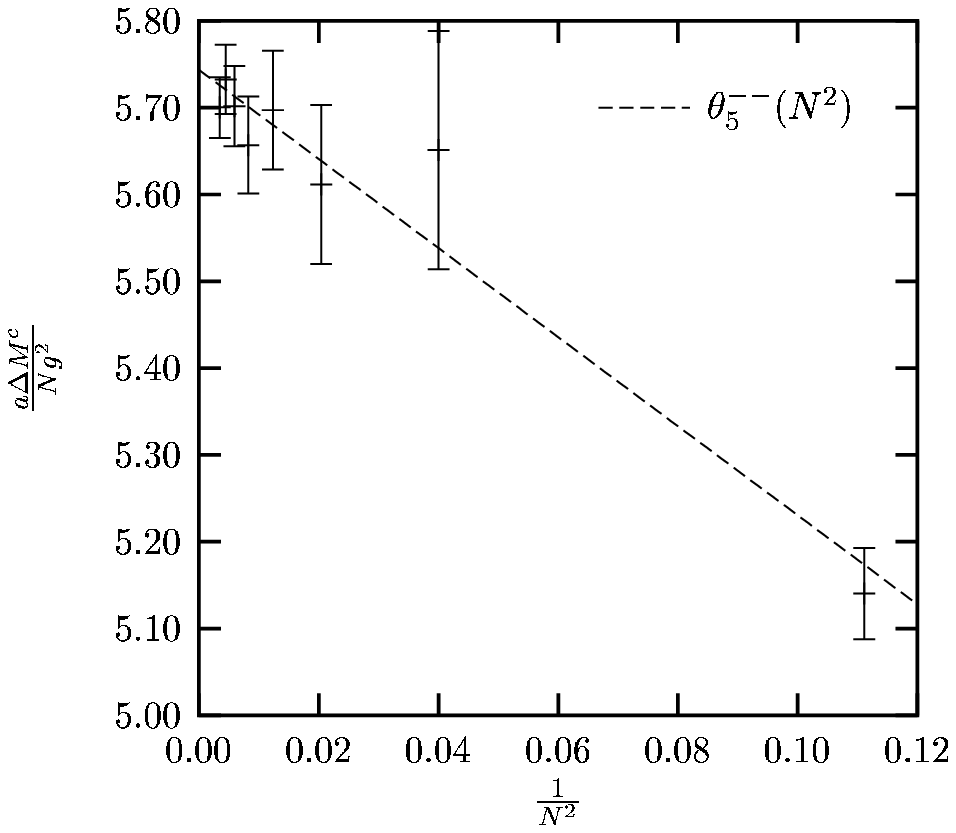}
                     }
\caption{The 2+1 dimensional continuum limit $2^{--}$ SU($N$) massgaps in units of $N
 g^2/a$ as functions of $1/N^2$. The dashed lines are fits given in
 \eqn{theta--}.} 
\label{as-largen-conv-spin2}  
\end{figure}

\begin{figure}
\centering
                       
\includegraphics[width=10cm]{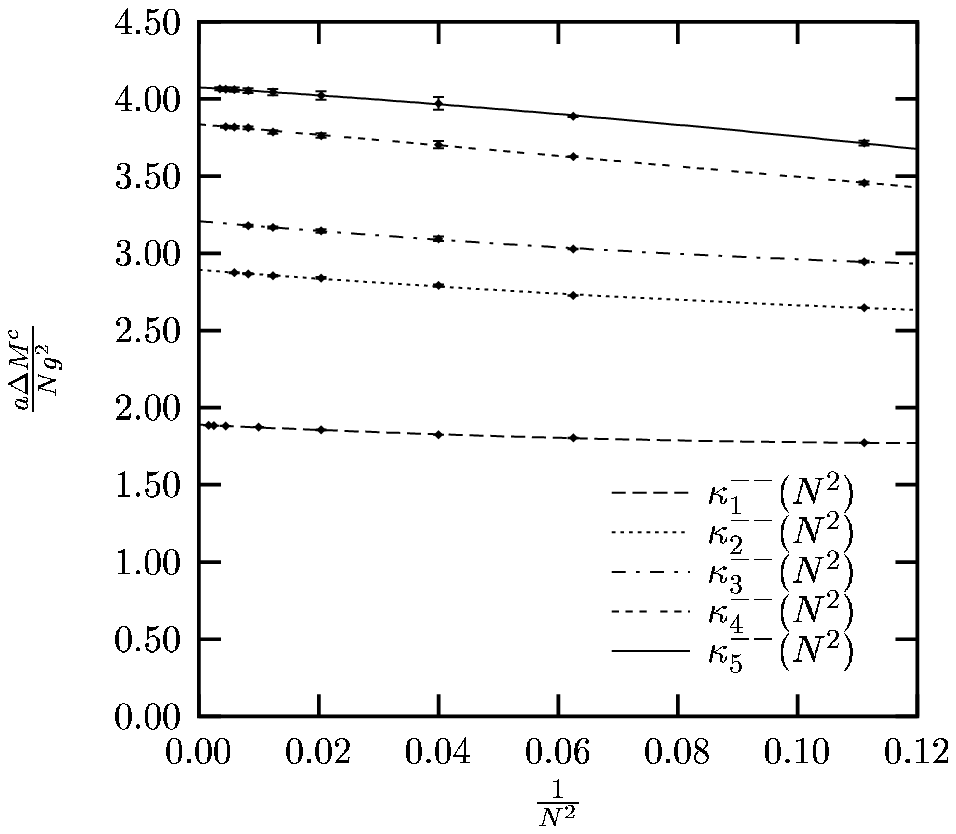}

\caption{An estimate of the continuum limit mass spectrum for 
$0^{--}$ SU($N$) glueballs in units
of $N g^2/a$ as a function of $1/N^2$. The lines are fits to the 
models of \eqn{kappa--}.}
\label{AS-spectrum}  
\end{figure}

\begin{figure}
\centering
                       
\includegraphics[width=10cm]{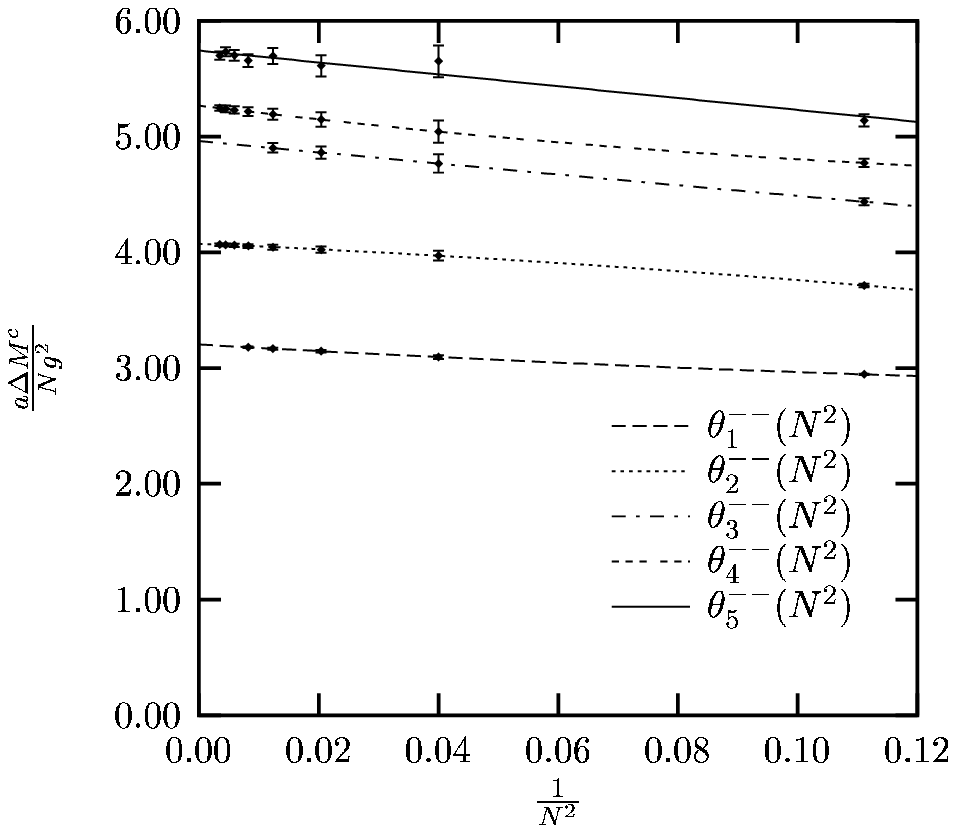}

\caption{An estimate of the continuum limit mass spectrum for $2^{--}$ SU($N$) glueballs in units
of $N g^2/a$ as a function of $1/N^2$. The lines are fits to the 
models of \eqn{theta--}.}
\label{AS-spin2-spectrum}  
\end{figure}

\subsection{An empirical observation}

\begin{figure}
\centering
                       
\includegraphics[width=10cm]{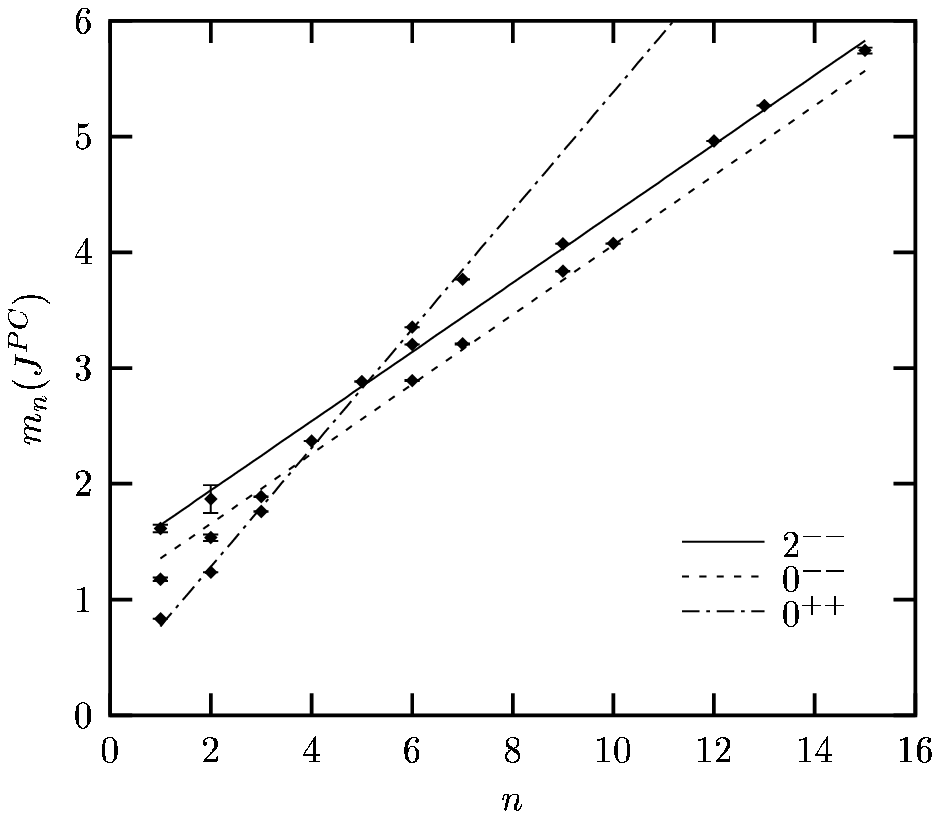}

\caption{A choice of enumeration of glueball masses and the fits of \eqn{spec-fits}.}
\label{spectra+model}  
\end{figure}

In this subsection we present an empirical observation that could
possibly be useful in the construction of simple glueball
models. Since Lucini and Teper include many small area Wilson loops in
their construction of excited states, we assume that they calculate
the lowest few glueball states without omission. We assume the
calculations presented here give higher mass glueball
states. Presumably, the correct enumeration of these states is not
possible without the inclusion of nonrectangular states in the
minimisation basis. For example, the third lowest mass state in a given sector calculated here could be the true seventh lowest 
mass state. However, when our results are combined with those of Lucini and Teper an interesting empirical observation can be made. We can choose a labelling of the $0^{++}$, $0^{--}$ and $2^{--}$ excited states such that the large $N$ limit of their masses  lie on a straight line as shown in \fig{spectra+model}. We have included the results of Lucini and Teper for $0^{--}$, $0^{--*}$, $2^{--}$ and  $2^{--*}$ which do not correspond to any of the states 
calculated here. The straight lines are fits to the model,
\bea
m_n(J^{PC}) = \gamma_1 (2 n + \gamma_2)  
\eea
where $\gamma_1$ and $\gamma_2$ are parameters and $\gamma_2$ is restricted to integer values. For the $J^{PC}$ states considered we obtain the following best fit models
\bea
m_n(0^{++}) &=& (0.256\pm 0.002)(2 n + 1) \nn\\
m_n(0^{--}) &=& (0.151\pm 0.002)(2 n + 5) \nn\\
m_n(2^{--}) &=& (0.1495\pm 0.0008)(2 n + 7).
\label{fitmodspec}
\eea
Here $n\ge 1$ labels the $n$-th lowest mass state. We note that the fit improves in accuracy as $n$ is increased. It is interesting to note that the slopes of the $0^{--}$ and $2^{--}$ models are consistent suggesting that the constant of proportionality does not depend on $J$. Another interesting observation can be made by recasting the models in the form 
\bea
m_n(0^{++}) &=& (0.256\pm 0.002)(2 n + 0 + 1) \nn\\
m_n(0^{--}) &=& (0.151\pm 0.002)(2 n + 4 + 1) \nn\\
m_n(2^{--}) &=& (0.1495\pm 0.0008)(2 n + 6+ 1).
\label{spec-fits}
\eea  
We notice a similarity with the two dimensional harmonic oscillator spectrum,
\bea
E_n \propto 2 n + J + 1,
\eea
when we take into consideration the ambiguity modulo 4 of spin identification on the lattice.
With this in mind we propose a simple model for the $J^{PC}$ spectrum
\bea
m_n(J^{PC}) = \zeta_{PC}\left[2n + \gamma(J^{PC}) + 1\right], 
\label{conjmod}
\eea 
where $\zeta_{PC}$ is a spin independent parameter and
$\gamma(J^{PC})$ is an integer for which $\gamma(J^{PC}) = J\, {\rm
mod}\, 4$. From \eqn{fitmodspec} we have $\zeta_{--} \approx 0.15$ and
$\zeta_{++} = 0.256\pm 0.002$. To check this simple model we can
attempt to predict the lowest lying states obtained by Lucini and
Teper in the sectors that have not been considered in this paper. We
start with the $2^{++}$ sector in which Teper and Lucini obtain
$1.359(12)$ and $1.822(62)$ for the $N\rightarrow \infty$ limit of the
$2^{++}$ and $2^{++*}$ glueball masses in units of $N g^2/a$. The
predictions of \eqn{conjmod}, with $\gamma(2^{++}) = 2$ and
$\zeta_{++} = 0.256$, are 1.28 and 1.79 for the 1st and 2nd
lowest glueball masses in units of $N g^2/a$. As we would expect the
prediction is better for the higher mass state. We can also consider
the $1^{++}$ sector which has an equivalent spectrum to $1^{-+}$ due
to the phenomenon of parity doubling~\cite{Teper:1998te}. Lucini and
Teper obtain $1.98(8)$ for the mass of the $1^{++}$ glueball in units
of $N g^2/a$. The model of \eqn{conjmod}, with $\gamma(1^{++})=5$ and
$\zeta_{++} = 0.256$ gives 2.048. In the $1^{--}$ sector the agreement
is not as good, with Lucini and Teper obtaining $1.85(13)$ for the
mass of the $1^{--}$ glueball and the model of \eqn{conjmod}, with
$\gamma(1^{--})= 9$ and $\zeta_{--}= 0.15$, giving 1.812. 

It will be interesting to see if the model suggested here stands up to
further calculations with an extended minimisation basis and in other
$J^{PC}$ sectors. The analytic techniques presented here are at a
great advantage to the standard Monte Carlo techniques of Lagrangian
LGT for the purpose of testing glueball models. High order excited
states are readily accessible; in a variational calculation with $s$
states in the minimisation basis, $s$ glueball states are
accessible. This is in contrast to competing Lagrangian calculations
in which only $3$ states are currently accessible in some $J^{PC}$ sectors.

\subsection{Discussion}

In this section we have calculated variational mass spectra for pure SU($N$)
gauge theory with a simple basis of rectangular states in 2+1
dimensions. 
Such a basis
is easy to work with computationally and is therefore a good
starting point. However to accurately explore the pure gauge 
mass spectra additional states need to be included. Perhaps a more
suitable basis would be the set of all states with an area less than some 
maximum value which would define the order of the calculation. 
The next stage beyond this would be to include
higher representation states in the form of Wilson loops in which some
links are covered more than once. In this way 
multiple glueball states could presumably be explored. 
The inclusion of additional small area states in the basis, in
particular states which are not symmetric about reflections in
coordinate axes, would also allow the calculation of spin $1$
glueball masses and also massgaps in the $P=-C$ sector. 


It is important to point out that the inclusion of additional states in
our minimisation basis does not present a significant challenge. The only
complication would be in counting the possible overlaps of particular
states. While more difficult than the counting required
here, the process could presumably be automated using symbolic programing and
techniques from graph theory. The analytic techniques used here would
still be applicable until higher representation states were included
in the minimisation basis. At
that stage further character integrals would be required to handle the 
additional integrals arising in the calculation.

A final extension of the method presented here would be to make use of
an improved ground state; a ground state which includes
additional Wilson loops in the exponent. However, the
advantage of using analytic techniques would then be lost, unless new technology
was developed for tackling the resulting integrals. One would need
to calculate the required matrix elements via Monte Carlo simulation
on small lattices, losing the advantage of working on an infinite
lattice with analytic expressions that we have here.

\section{Conclusion}

In this paper we have applied the analytic techniques developed in
\rcite{thesis-paper1}, in a study of the large $N$ glueball
mass spectrum in 2+1 dimensions.


In \sect{extrapolation} we calculated glueball masses at finite
$N$, with $N$ as large as 25 in some cases. This allowed accurate $N\rightarrow
\infty$ extrapolations to be made. Evidence of leading 
order $1/N^2$ finite $N$ corrections to the glueball masses was
obtained, confirming a specific prediction of large $N$ gauge theory.
The interpretation of the possible scaling regions was discussed and
agreement with the Lagrangian study of Lucini and Teper was obtained
in some cases in the $N\rightarrow \infty$ limit. The discrepancies
are attributable to our use of only rectangular states in the
minimisation basis. Without the inclusion of small
area, nonrectangular states, it is likely that some low energy states are inaccessible. 
Further work will fill in the incomplete
spectra presented here. Interesting empirical observations were made
when the results presented here were combined with those of Lucini and
Teper; the enumeration of excited states can be chosen so that the glueball 
mass spectrum has the structure
of a two dimensional harmonic oscillator. To develop this observation
into a model would require a more complete calculation of the mass
spectrum with nonrectangular states in the minimisation basis and
possibly a more complicated vacuum trial state.

Clearly much work remains to be done before an accuruate picture of the
2+1 dimensional pure SU($N$) gauge theory spectrum is achieved within a
Hamiltonian variational approach. We have demonstrated however that
the analytic techniques of Hamiltonian LGT can be used for massgap calculations with $N$ as large as 25 on a desktop computer. 
This is significantly closer to
the $N\rightarrow \infty$ limit than is currently possible with the
Lagrangian approach using supercomputers. The inclusion of additional
states in the minimisation basis, while not presenting a major
challenge, will allow a thorough study of the 2+1 dimensional 
pure SU($N$) mass
spectrum at least up to $N=25$. 

\begin{acknowledgments}
We wish to acknowledge useful and interesting discussions with J.~A.~L.~McIntosh and L.~C.~L.~Hollenberg.
\end{acknowledgments}

\bibliography{thesis-paper2}

\begin{thebibliography}{28}
\expandafter\ifx\csname natexlab\endcsname\relax\def\natexlab#1{#1}\fi
\expandafter\ifx\csname bibnamefont\endcsname\relax
  \def\bibnamefont#1{#1}\fi
\expandafter\ifx\csname bibfnamefont\endcsname\relax
  \def\bibfnamefont#1{#1}\fi
\expandafter\ifx\csname citenamefont\endcsname\relax
  \def\citenamefont#1{#1}\fi
\expandafter\ifx\csname url\endcsname\relax
  \def\url#1{\texttt{#1}}\fi
\expandafter\ifx\csname urlprefix\endcsname\relax\def\urlprefix{URL }\fi
\providecommand{\bibinfo}[2]{#2}
\providecommand{\eprint}[2][]{\url{#2}}

\bibitem[{\citenamefont{Teper}(1999)}]{Teper:1998te}
\bibinfo{author}{\bibfnamefont{M.~J.} \bibnamefont{Teper}},
  \bibinfo{journal}{Phys. Rev.} \textbf{\bibinfo{volume}{D59}},
  \bibinfo{pages}{014512} (\bibinfo{year}{1999}),
  \eprint[http://arXiv.org/abs]{hep-lat/9804008}.

\bibitem[{\citenamefont{Lucini and Teper}(2002)}]{Lucini:2002wg}
\bibinfo{author}{\bibfnamefont{B.}~\bibnamefont{Lucini}} \bibnamefont{and}
  \bibinfo{author}{\bibfnamefont{M.}~\bibnamefont{Teper}}
  (\bibinfo{year}{2002}), \eprint[http://arXiv.org/abs]{hep-lat/0206027}.

\bibitem[{\citenamefont{Lucini and Teper}(2001)}]{Lucini:2001ej}
\bibinfo{author}{\bibfnamefont{B.}~\bibnamefont{Lucini}} \bibnamefont{and}
  \bibinfo{author}{\bibfnamefont{M.}~\bibnamefont{Teper}},
  \bibinfo{journal}{JHEP} \textbf{\bibinfo{volume}{06}}, \bibinfo{pages}{050}
  (\bibinfo{year}{2001}), \eprint[http://arXiv.org/abs]{hep-lat/0103027}.

\bibitem[{\citenamefont{Carlsson and McKellar}(2003)}]{thesis-paper1}
\bibinfo{author}{\bibfnamefont{J.}~\bibnamefont{Carlsson}} \bibnamefont{and}
  \bibinfo{author}{\bibfnamefont{B.~H.~J.} \bibnamefont{McKellar}}
  (\bibinfo{year}{2003}), \eprint{hep-lat/0303016}.

\bibitem[{\citenamefont{'t~Hooft}(1974)}]{'tHooft:1974jz}
\bibinfo{author}{\bibfnamefont{G.}~\bibnamefont{'t~Hooft}},
  \bibinfo{journal}{Nucl. Phys.} \textbf{\bibinfo{volume}{B72}},
  \bibinfo{pages}{461} (\bibinfo{year}{1974}).

\bibitem[{\citenamefont{Witten}(1979)}]{Witten:1979kh}
\bibinfo{author}{\bibfnamefont{E.}~\bibnamefont{Witten}},
  \bibinfo{journal}{Nucl. Phys.} \textbf{\bibinfo{volume}{B160}},
  \bibinfo{pages}{57} (\bibinfo{year}{1979}).

\bibitem[{\citenamefont{Eguchi and Kawai}(1982)}]{Eguchi:1982nm}
\bibinfo{author}{\bibfnamefont{T.}~\bibnamefont{Eguchi}} \bibnamefont{and}
  \bibinfo{author}{\bibfnamefont{H.}~\bibnamefont{Kawai}},
  \bibinfo{journal}{Phys. Rev. Lett.} \textbf{\bibinfo{volume}{48}},
  \bibinfo{pages}{1063} (\bibinfo{year}{1982}).

\bibitem[{\citenamefont{Maldacena}(1998)}]{Maldacena:1998re}
\bibinfo{author}{\bibfnamefont{J.~M.} \bibnamefont{Maldacena}},
  \bibinfo{journal}{Adv. Theor. Math. Phys.} \textbf{\bibinfo{volume}{2}},
  \bibinfo{pages}{231} (\bibinfo{year}{1998}),
  \eprint[http://arXiv.org/abs]{hep-th/9711200}.

\bibitem[{\citenamefont{Witten}(1998)}]{Witten:1998qj}
\bibinfo{author}{\bibfnamefont{E.}~\bibnamefont{Witten}},
  \bibinfo{journal}{Adv. Theor. Math. Phys.} \textbf{\bibinfo{volume}{2}},
  \bibinfo{pages}{253} (\bibinfo{year}{1998}),
  \eprint[http://arXiv.org/abs]{hep-th/9802150}.

\bibitem[{\citenamefont{Csaki et~al.}(1999{\natexlab{a}})\citenamefont{Csaki,
  Ooguri, Oz, and Terning}}]{Csaki:1998qr}
\bibinfo{author}{\bibfnamefont{C.}~\bibnamefont{Csaki}},
  \bibinfo{author}{\bibfnamefont{H.}~\bibnamefont{Ooguri}},
  \bibinfo{author}{\bibfnamefont{Y.}~\bibnamefont{Oz}}, \bibnamefont{and}
  \bibinfo{author}{\bibfnamefont{J.}~\bibnamefont{Terning}},
  \bibinfo{journal}{JHEP} \textbf{\bibinfo{volume}{01}}, \bibinfo{pages}{017}
  (\bibinfo{year}{1999}{\natexlab{a}}),
  \eprint[http://arXiv.org/abs]{hep-th/9806021}.

\bibitem[{\citenamefont{de~Mello~Koch et~al.}(1998)\citenamefont{de~Mello~Koch,
  Jevicki, Mihailescu, and Nunes}}]{deMelloKoch:1998qs}
\bibinfo{author}{\bibfnamefont{R.}~\bibnamefont{de~Mello~Koch}},
  \bibinfo{author}{\bibfnamefont{A.}~\bibnamefont{Jevicki}},
  \bibinfo{author}{\bibfnamefont{M.}~\bibnamefont{Mihailescu}},
  \bibnamefont{and} \bibinfo{author}{\bibfnamefont{J.~P.} \bibnamefont{Nunes}},
  \bibinfo{journal}{Phys. Rev.} \textbf{\bibinfo{volume}{D58}},
  \bibinfo{pages}{105009} (\bibinfo{year}{1998}),
  \eprint[http://arXiv.org/abs]{hep-th/9806125}.

\bibitem[{\citenamefont{Zyskin}(1998)}]{Zyskin:1998tg}
\bibinfo{author}{\bibfnamefont{M.}~\bibnamefont{Zyskin}},
  \bibinfo{journal}{Phys. Lett.} \textbf{\bibinfo{volume}{B439}},
  \bibinfo{pages}{373} (\bibinfo{year}{1998}),
  \eprint[http://arXiv.org/abs]{hep-th/9806128}.

\bibitem[{\citenamefont{Brower et~al.}(2000{\natexlab{a}})\citenamefont{Brower,
  Mathur, and Tan}}]{Brower:1999nj}
\bibinfo{author}{\bibfnamefont{R.~C.} \bibnamefont{Brower}},
  \bibinfo{author}{\bibfnamefont{S.~D.} \bibnamefont{Mathur}},
  \bibnamefont{and} \bibinfo{author}{\bibfnamefont{C.-I.} \bibnamefont{Tan}},
  \bibinfo{journal}{Nucl. Phys.} \textbf{\bibinfo{volume}{B574}},
  \bibinfo{pages}{219} (\bibinfo{year}{2000}{\natexlab{a}}),
  \eprint[http://arXiv.org/abs]{hep-th/9908196}.

\bibitem[{\citenamefont{Brower et~al.}(2000{\natexlab{b}})\citenamefont{Brower,
  Mathur, and Tan}}]{Brower:2000rp}
\bibinfo{author}{\bibfnamefont{R.~C.} \bibnamefont{Brower}},
  \bibinfo{author}{\bibfnamefont{S.~D.} \bibnamefont{Mathur}},
  \bibnamefont{and} \bibinfo{author}{\bibfnamefont{C.-I.} \bibnamefont{Tan}},
  \bibinfo{journal}{Nucl. Phys.} \textbf{\bibinfo{volume}{B587}},
  \bibinfo{pages}{249} (\bibinfo{year}{2000}{\natexlab{b}}),
  \eprint[http://arXiv.org/abs]{hep-th/0003115}.

\bibitem[{\citenamefont{Russo}(1999)}]{Russo:1998mm}
\bibinfo{author}{\bibfnamefont{J.~G.} \bibnamefont{Russo}},
  \bibinfo{journal}{Nucl. Phys.} \textbf{\bibinfo{volume}{B543}},
  \bibinfo{pages}{183} (\bibinfo{year}{1999}),
  \eprint[http://arXiv.org/abs]{hep-th/9808117}.

\bibitem[{\citenamefont{Csaki et~al.}(1999{\natexlab{b}})\citenamefont{Csaki,
  Russo, Sfetsos, and Terning}}]{Csaki:1999vb}
\bibinfo{author}{\bibfnamefont{C.}~\bibnamefont{Csaki}},
  \bibinfo{author}{\bibfnamefont{J.}~\bibnamefont{Russo}},
  \bibinfo{author}{\bibfnamefont{K.}~\bibnamefont{Sfetsos}}, \bibnamefont{and}
  \bibinfo{author}{\bibfnamefont{J.}~\bibnamefont{Terning}},
  \bibinfo{journal}{Phys. Rev.} \textbf{\bibinfo{volume}{D60}},
  \bibinfo{pages}{044001} (\bibinfo{year}{1999}{\natexlab{b}}),
  \eprint[http://arXiv.org/abs]{hep-th/9902067}.

\bibitem[{\citenamefont{Dalley and van~de Sande}(2001)}]{Dalley:2000ye}
\bibinfo{author}{\bibfnamefont{S.}~\bibnamefont{Dalley}} \bibnamefont{and}
  \bibinfo{author}{\bibfnamefont{B.}~\bibnamefont{van~de Sande}},
  \bibinfo{journal}{Phys. Rev.} \textbf{\bibinfo{volume}{D63}},
  \bibinfo{pages}{076004} (\bibinfo{year}{2001}),
  \eprint[http://arXiv.org/abs]{hep-lat/0010082}.

\bibitem[{\citenamefont{Karabali and Nair}(1996)}]{Karabali:1996je}
\bibinfo{author}{\bibfnamefont{D.}~\bibnamefont{Karabali}} \bibnamefont{and}
  \bibinfo{author}{\bibfnamefont{V.~P.} \bibnamefont{Nair}},
  \bibinfo{journal}{Phys. Lett.} \textbf{\bibinfo{volume}{B379}},
  \bibinfo{pages}{141} (\bibinfo{year}{1996}),
  \eprint[http://arXiv.org/abs]{hep-th/9602155}.

\bibitem[{\citenamefont{Karabali
  et~al.}(1998{\natexlab{a}})\citenamefont{Karabali, Kim, and
  Nair}}]{Karabali:1998wk}
\bibinfo{author}{\bibfnamefont{D.}~\bibnamefont{Karabali}},
  \bibinfo{author}{\bibfnamefont{C.-J.} \bibnamefont{Kim}}, \bibnamefont{and}
  \bibinfo{author}{\bibfnamefont{V.~P.} \bibnamefont{Nair}},
  \bibinfo{journal}{Nucl. Phys.} \textbf{\bibinfo{volume}{B524}},
  \bibinfo{pages}{661} (\bibinfo{year}{1998}{\natexlab{a}}),
  \eprint[http://arXiv.org/abs]{hep-th/9705087}.

\bibitem[{\citenamefont{Karabali
  et~al.}(1998{\natexlab{b}})\citenamefont{Karabali, Kim, and
  Nair}}]{Karabali:1998yq}
\bibinfo{author}{\bibfnamefont{D.}~\bibnamefont{Karabali}},
  \bibinfo{author}{\bibfnamefont{C.-J.} \bibnamefont{Kim}}, \bibnamefont{and}
  \bibinfo{author}{\bibfnamefont{V.~P.} \bibnamefont{Nair}},
  \bibinfo{journal}{Phys. Lett.} \textbf{\bibinfo{volume}{B434}},
  \bibinfo{pages}{103} (\bibinfo{year}{1998}{\natexlab{b}}),
  \eprint[http://arXiv.org/abs]{hep-th/9804132}.

\bibitem[{\citenamefont{Nair}(2002)}]{Nair:2002uj}
\bibinfo{author}{\bibfnamefont{V.~P.} \bibnamefont{Nair}},
  \bibinfo{journal}{Nucl. Phys. Proc. Suppl.} \textbf{\bibinfo{volume}{108}},
  \bibinfo{pages}{194} (\bibinfo{year}{2002}).

\bibitem[{\citenamefont{Carlsson and McKellar}(2001)}]{Carlsson:2001wp}
\bibinfo{author}{\bibfnamefont{J.}~\bibnamefont{Carlsson}} \bibnamefont{and}
  \bibinfo{author}{\bibfnamefont{B.~H.~J.} \bibnamefont{McKellar}},
  \bibinfo{journal}{Phys. Rev.} \textbf{\bibinfo{volume}{D64}},
  \bibinfo{pages}{094503} (\bibinfo{year}{2001}), \eprint{hep-lat/0105018}.

\bibitem[{\citenamefont{Kogut and Susskind}(1975)}]{Kogut:1975ag}
\bibinfo{author}{\bibfnamefont{J.~B.} \bibnamefont{Kogut}} \bibnamefont{and}
  \bibinfo{author}{\bibfnamefont{L.}~\bibnamefont{Susskind}},
  \bibinfo{journal}{Phys. Rev.} \textbf{\bibinfo{volume}{D11}},
  \bibinfo{pages}{395} (\bibinfo{year}{1975}).

\bibitem[{\citenamefont{Fang et~al.}(2002)\citenamefont{Fang, Hui, Chen, and
  Sch{\"u}tte}}]{Fang:2002ps}
\bibinfo{author}{\bibfnamefont{X.-Y.} \bibnamefont{Fang}},
  \bibinfo{author}{\bibfnamefont{P.}~\bibnamefont{Hui}},
  \bibinfo{author}{\bibfnamefont{Q.-Z.} \bibnamefont{Chen}}, \bibnamefont{and}
  \bibinfo{author}{\bibfnamefont{D.}~\bibnamefont{Sch{\"u}tte}},
  \bibinfo{journal}{Phys. Rev.} \textbf{\bibinfo{volume}{D65}},
  \bibinfo{pages}{114505} (\bibinfo{year}{2002}).

\bibitem[{\citenamefont{Chen et~al.}(1995)\citenamefont{Chen, Luo, Guo, and
  Fang}}]{Chen:1995ca}
\bibinfo{author}{\bibfnamefont{Q.-Z.} \bibnamefont{Chen}},
  \bibinfo{author}{\bibfnamefont{X.-Q.} \bibnamefont{Luo}},
  \bibinfo{author}{\bibfnamefont{S.-H.} \bibnamefont{Guo}}, \bibnamefont{and}
  \bibinfo{author}{\bibfnamefont{X.-Y.} \bibnamefont{Fang}},
  \bibinfo{journal}{Phys. Lett.} \textbf{\bibinfo{volume}{B348}},
  \bibinfo{pages}{560} (\bibinfo{year}{1995}),
  \eprint[http://arXiv.org/abs]{hep-ph/9502235}.

\bibitem[{\citenamefont{Leonard}(2001)}]{ConradPhD}
\bibinfo{author}{\bibfnamefont{C.~R.} \bibnamefont{Leonard}}, Ph.D. thesis,
  \bibinfo{school}{The University of Melbourne} (\bibinfo{year}{2001}).

\bibitem[{\citenamefont{Hamer}(1996)}]{Hamer:1996ub}
\bibinfo{author}{\bibfnamefont{C.~J.} \bibnamefont{Hamer}},
  \bibinfo{journal}{Phys. Rev.} \textbf{\bibinfo{volume}{D53}},
  \bibinfo{pages}{7316} (\bibinfo{year}{1996}).

\bibitem[{\citenamefont{Wichmann}(2001)}]{WichmannPhD}
\bibinfo{author}{\bibfnamefont{A.}~\bibnamefont{Wichmann}}, Ph.D. thesis,
  \bibinfo{school}{ITKP Bonn} (\bibinfo{year}{2001}).

\end{thebibliography}
\end{document}